\documentclass[journal]{new-aiaa}

\usepackage{graphicx}      %
\usepackage{natbib}        %
\usepackage{xcolor}
\usepackage{graphicx}
\usepackage{amsmath}

\usepackage{amsthm}
\usepackage[version=4]{mhchem}
\usepackage{longtable,tabularx}
\usepackage{bm}
\usepackage{float}
\usepackage{algorithm}
\usepackage{algcompatible}
\usepackage{multirow}
\usepackage{subcaption}
\usepackage{textcase}
\usepackage[normalem]{ulem}
\usepackage{mathtools}

\newcommand{\multiline}[1]{%
  \begin{tabularx}{\dimexpr\linewidth-\ALG@thistlm}[t]{@{}X@{}}
    #1
  \end{tabularx}
}

\newcommand{\range}[1]{0, \dots, #1}
\newcommand{\rangeset}[1]{\{\range{#1}\}}

\newcommand{\T}[0]{{\rm T}}

\newtheorem{lem}{Lemma}
\newtheorem{theorem}{Theorem}
\newtheorem{remark}{Remark}
\newtheorem{cor}{Corollary}
\newtheorem{defin}{Definition}

\title{Occlusion-Aware Ground Target Tracking by a Dubins Vehicle using Visibility Volumes 
}
\author{Collin Hague\footnote{Graduate Student, Department of Mechanical Engineering and Engineering Science.} and Artur Wolek\footnote{Assistant Professor, Department of Mechanical Engineering and Engineering Science, AIAA Senior Member.}}
\affil{
University of North Carolina at Charlotte, Charlotte, North Carolina, 28223}

\begin{document}
\maketitle

\begin{abstract}
This paper considers the problem of tracking a point of interest (POI) moving along a known trajectory on the ground with an uncrewed aerial vehicle (UAV) modeled as a Dubins vehicle using a line-of-sight (LOS) sensor through an urban environment that may occlude the POI. A visibility volume (VV) encodes a time-varying, three-dimensional representation of the sensing constraints for a particular POI position. A constant-altitude, translating, and radially time-varying circular standoff orbit is then inscribed within the dynamically changing VV centered at the POI position. The time-varying VV is approximated by placing static VVs along the POI's trajectory using an adaptive metric that restricts the volume change of consecutive VVs to below a specified rate. The time-varying circular standoff orbit is proven to be feasible for a Dubins vehicle and approximated with a piecewise set of linearly interpolated circular orbits inside the static VVs. A steering controller is derived that drives the UAV to the time-varying standoff orbit. Numerical simulations and a flight test illustrate the proposed approach.
\end{abstract}

\section{Nomenclature}
{\renewcommand\arraystretch{1.0}
\noindent\begin{longtable*}{@{}l @{\quad=\quad} l@{}}
$(x, y, z)$ & inertial directions (m) easting, northing, and up\\
$\{{\bm e}_x, {\bm e}_y, {\bm e}_z \}$  & inertial direction vectors (east, north, up) \\
$\psi$ & yaw angle (rad) about $z$ direction\\
$t$ & time (s)\\
$\bm{q}$ & configuration of the UAV; $[x, y, \psi]^{\rm T} \in {\rm SE}(2)$\\
$v$ & forward velocity of the UAV (m/s)\\
$u_{\psi}$ & turn-rate control for the UAV (rad/s)\\
$\kappa$, $\kappa_s$ & signed and unsigned curvature of a trajectory (1/m)\\
$r_{\rm min}$ & minimum turn radius of the UAV (m); $1 / \kappa_{\rm max}$\\
$h_{\rm UAV}$ & flight altitude of the UAV (m)\\
$B$ & set of $N_B$ 2.5-dimensional objects; $\{B_0,\ldots,B_{N_B -1}\}$ \\
$A_i$ & the order set of points representing the base of the $i$th object $B_i$ \\
$h_i$ & the height of the $i$th object $B_i$ (m)\\
$\emptyset$ & an empty set\\
$\partial A_i$ & polygon with boundary $A_i$\\
${\rm int}(A_i)$ & region on the interior of $A_i$ \\
$G$ & graph consisting of nodes $N$ and edges $E$ \\
$\bm{g}$ & inertial position of the moving point of interest (m); $[g_x, g_y]^{\rm T}$ \\ 
$v_{\rm g}$ & magnitude of the velocity (m/s) for the moving point of interest\\
$\gamma$ & direction (rad) of the point of interest's velocity\\
$\mathcal{P}$ & set of points sampled along the ground plane\\
$l(\bm{p}_i,\bm{p}_j)$ & straight line between $\bm{p}_i$ and $\bm{p}_j$ along the ground plane\\
$d_{\rm max}$ & maximum viewing distance for imaging (m)\\
$\alpha(\bm{g})$, $\hat{\alpha}(\bm{g})$ & visibility volume and its approximation for target position $\bm{g}$ \\
$h_{\rm building}$ & maximum height of the tallest building (m) \\
$h_{\rm feasible}$ & maximum height of the feasible airspace (m) \\
$F$ & the feasible airspace for the UAV\\
$L(\tau;\bm{g}, \bm{\rho})$ & equation of a line from  $\bm{g}$ to $\bm{\rho}$ parameterized by $\tau$ in $\mathbb{R}^3$\\
$\theta$ & angle (rad) about the point $\bm{g}$ measured from a line parallel to the $x$ axis\\
$O(\theta, t)$ & visibility orbit\\
$R(t)$ & radius of a morphing visibility orbit (m)\\
$\bm{x}_\mathcal{I}$ & inertial trajectory (m) of the UAV in the $xy$ plane; $\bm{x}_\mathcal{I} \in \mathbb{R}^2$\\
$\bm{e}_r$,$\bm{e}_\theta$ & a polar coordinate system from the point $\bm{g}$\\
$r$ & UAV's distance (m) from the POI\\
$d(\hat{\alpha}(\bm{g}_i), \hat{\alpha}(\bm{g}_j))$ & measure (m$^3$) of the difference between two triangular meshes approximating visibility volumes\\
$d_{\rm cutoff}$ & maximum difference (m$^3$) between two visibility volumes\\
$V_1$, $V_2$ & Lyapunov-like scalar functions\\
$a_1$ & a function that is only zero on $O(\theta, t)$\\
$\Phi$ & nonnegative function scaling the attraction to the orbit $O(\theta, t)$ \\
$H$ & strictly positive or negative function scaling the circulation about the orbit $O(\theta, t)$ \\
$\bm{\xi}$ & inertial coordinates (m) for a kinematic particle; $\xi \in \mathbb{R}^2$ \\
$\bm{u}$ & a vector field (m/s) that forces a particle onto an orbit $O(\theta, t)$; $\bm{u} \in \mathbb{R}^2$ \\
$\bm{M}$ & a matrix component of the vector field $\bm{u}$\\
$\dot{\bm{\sigma}}$ & a vector $[\partial a_1/\partial t, 0]^{\rm T}$\\
$\beta$ & circulation parameter \\
$\psi_d$ & desired heading angle (rad) \\
$\Phi'$ & the scalar component of $\Phi\nabla V_1$\\
$k_\psi$ & positive proportional gain for heading controller\\
\end{longtable*}}

\section{Introduction}

\lettrine{U}{ncrewed} aerial vehicles (UAVs) are widely used for airborne sensing tasks that require maintaining a direct line of sight (LOS) to a target point of interest (POI). Applications of tracking a moving POI include cinematography, filming sporting events and parades, search and rescue, and police and military surveillance \cite{Zhang2023Auto, schedl2021autonomous}. 
For UAV tracking missions where a POI moves along a road through an urban environment, the directional sensor's view of a POI may be obstructed by buildings, tunnels, bridges, trees, and other structures.
This work creates a planning and control method to ensure a UAV maintains an unobstructed LOS to the a POI, moving along a known trajectory, while continuously orbiting it. By orbiting the POI, the UAV can observe it from various angles. Moreover, an encircling motion is required when the UAV cannot hover over the POI (e.g., if the POI's speed is less than the UAV's minimum speed).
The POI moves over a road network represented as a graph with nodes representing points along roads and edges representing road segments joining two nodes.
To calculate the time-varying sensing constraints of the POI, a time-varying visibility volume (VV) captures airspace, sensing quality, and LOS constraints.
As the POI moves and encounters new building geometry, its VV changes shape. To capture the variation of the VV, the road edges are sampled to form a space (and time) parameterized sequence of VVs. VV creation is described in our prior work \cite{Willis2024, Hague2023}. To balance the computational load of generating VVs and capturing changes in VV geometry, a bisection-inspired adaptive discretization method measures the similarity between two VVs, placing another VV in between if the geometry differs more than a specified rate. In this work, a UAV is approximated by a Dubins vehicle \cite{dubins1957curves} operating at a constant altitude. The VVs are sliced at the UAV's operating altitude, creating visibility polygons (VPs). Circles are inscribed in the VPs to produce static visibility orbits (VOs). Lastly, the time-varying VO is constructed by linearly interpolating neighboring static VOs.  
If the UAV remains on the time-varying VO, the UAV is inside the POI's VV. To drive the UAV to converge to this time-varying VO, a vector field provides a reference heading to a steering controller for the Dubins vehicle model.

The POI's environment determines how a UAV can position itself to view a POI. Many prior works consider environments that do not contain obstacles that may obscure a UAV's visual sensors. For example, some authors assume that both the UAV and target move in a two-dimensional (2D) plane 
\cite{Pothen2017, Lin2022, Srinivasu2022}, the UAV moves in three-dimensions (3D) while the target is constrained to a horizontal plane 
\cite{Zhang2023Auto, Anand2023}, or both the UAV and target move in 3D
\cite{Kim2021, Penin2018}.
Previous target tracking algorithms consider a target POI that moves over 
a discretized road network or group of cells \cite{Skoglar2012, Wolek.RAL.2020, Cook2014}.
In \cite{Watanabe2010}, the UAV must avoid colliding with the urban environment, but the tracking method does not consider the obstruction of a visual sensor. In \cite{Wolek.RAL.2020}, the target POI is constrained to an urban road network, but occlusions are not considered. Researchers have also explored target tracking with more general non-building obstacles. In \cite{Hausman2016}, a target moves in 2D and a team of multirotors moves in 3D while maneuvering to avoid obscuring another team member's view of the target. Reference \cite{Penin2018} considers spheres floating above the ground that UAVs must avoid and look beyond to track a target moving in 3D. The effect of mountainous terrain on target velocity is explored in \cite{Kim2021}. Finally, some studies consider buildings and other obstacles blocking visual sensors prohibiting measurement of the target \cite{Skoglar2012, Cook2014, Bhagat2020, Tyagi2021}---the POI is blocked if there is an obstructed line from the camera to the target.

Multirotors and fixed-wing UAVs are commonly used in target-tracking applications. Multirotor UAVs can be modeled as single integrators
\cite{Hausman2016}
or double integrators
\cite{Watanabe2010, Wolek.RAL.2020}. More complex rigid-body dynamics have also been considered 
\cite{Zhang2023Auto}
and are necessary when the vehicle's attitude determines the onboard camera's orientation.
Fixed-wing UAVs are often modeled with turn rate constraints, such as the acceleration-limited double integrator
\cite{Sinha2022}, Dubins car
\cite{Skoglar2012, Cook2014, Tyagi2021, Lin2022, Srinivasu2022, Anand2023}, and 3D Dubins airplane
\cite{Tyagi2021}. More sophisticated fixed-wing models are used when variable speed 
\cite{Hu2021}
 or roll angle \cite{Mali2020} are relevant when an aircraft can change elevation to look past obstacles \cite{WangJ2022}.

Methods for steering a vehicle to track a moving target can be grouped into (i) path-planning methods and (ii) feedback-control strategies. Path-planning methods explicitly plan a path over a fixed or moving time horizon to maintain the target in view. For example, \cite{Watanabe2010, Skoglar2012} investigated a single-step look-ahead strategy.
The optimal path for a fixed-wing UAV performing standoff tracking can be found with
model predictive control (MPC). MPC-based methods have also been considered for multirotors \cite{Hausman2016, Penin2018} and fixed-wing aircraft with
fixed cameras
\cite{Mali2020}, and gimballed cameras \cite{Tyagi2021, Hu2021, WangJ2022}.
In \cite{Zhang2023Auto} 
$\rm A^*$ planning is used for tracking a target with a known path. 
Feedback-control strategies define a control objective, maintaining a target-relative position or circulating a target, by employing error feedback.
One example is a Lyapunov vector field guidance law that steers fixed-wing UAVs onto fixed-radius-circular orbits around a target with constant velocity motion models
\cite{Anand2023}, acceleration limited double integrator models
\cite{Sinha2022}, or single integrator models with a curvature constraint \cite{Pothen2017}.
Other methods of forcing circulation on fixed-radius-circular orbits are sliding mode control---for example, driving a fixed-wing to circle a target \cite{Lin2022}---and other nonlinear control laws that force circulation
\cite{Srinivasu2022}. Feedback control laws have also been proposed to steer fixed-wing UAVs
\cite{Skoglar2012}, and multirotors
\cite{Wolek.RAL.2020}
to positions relative to the target.
Reinforcement learning is used to train a neural network to track a target in
\cite{Bhagat2020}.

This contributions of this paper are: (i) A computationally efficient method of designing time-varying circular reference orbits that are proven to be feasible for a Dubins vehicle and guarantee visibility to a moving POI in a geometrically complex environment; (ii) A guidance vector field to provide a reference heading to converge to a moving circular reference orbit with a time-varying radius; and (iii) A Lyapunov-based steering controller that forces the UAV onto the guidance vector field. A strategy for tuning the controller to satisfy the maximum turn-rate constraint is also discussed. The proposed tracking control and path planning techniques are demonstrated through numerical simulations in a complex urban environment and compared to methods that orbit the POI with a constant radius. A flight test illustrates implementation on a small multirotor UAV tracking a virtual POI on a road segment of a university campus.

The remainder of this paper is organized as follows. 
Section~\ref{sec:problem} explains the environment model, vehicle motion model, point of interest motion, and measurement model. The two problem statements in this work, reference orbit design and controller design, are given. Section~\ref{sec:polygons} finds the feasibility of the visibility orbit for a Dubins vehicle and describes an algorithm to adaptively place VVs along a road network to satisfy a limit of geometry change. Another proposed algorithm constructs the time-varying orbit inside the time-varying VPs. Section~\ref{sec:feed_controller} derives the guidance vector field and the steering controller that force the UAV onto the VO. Section~\ref{sec:results} analyzes the numerical simulations illustrating the proposed approach. Section~\ref{sec:flight_test} presents a flight test experiment. Section~\ref{sec:conclusion} concludes the paper.

\section{Problem Formulation}
\label{sec:problem}

This section formulates the problem of tracking a moving target through an urban environment. 
The UAV's motion is approximated by a Dubins vehicle, and its environment consists of 2.5D obstacles and a road network. The section introduces the POI's motion, the UAV's sensor model, and measurement of the POI. Lastly, the problem statements are formulated as creating the time-varying VO and designing a controller to attract the UAV to the time-varying VO.

\subsection{Vehicle Motion}
\label{sec:veh_model}
Consider a Dubins car model \cite{dubins1957curves} as an approximation of a UAV flying at a fixed altitude $h_{\rm UAV}$,
\begin{align}
    \dot{{\bm{q}}} &=  {\bm{f}}({\bm{q}}, u_\psi)\notag \\
    \begin{bmatrix}
        \dot{q}_x \\ \dot{q}_y \\ \dot{q}_\psi
    \end{bmatrix}
    &=
    \begin{bmatrix}
        v \cos{q_\psi} \\
        v \sin{q_\psi} \\
        u_{\psi}
    \end{bmatrix} ,
    \label{eq:dynamics}
\end{align}
where the state is $\bm{q} \in {\rm SE}(2)$, the position is $(q_x,q_y)$, the heading angle $q_\psi$ is measured counter-clockwise from the inertial positive $x$-axis, and the forward velocity $v$ is a constant value. 
The turn-rate control is bounded by 
\begin{equation}
     |u_{\psi}| \leq u_{\psi, {\rm max}}\label{eq:dynamics_constraint}\;,
\end{equation}
where $u_{\psi, {\rm max}}$ is the maximum allowable turn rate.
The bounded turn rate creates a minimum turn radius $r_{\rm min} = v / u_{\psi, {\rm max}}$ and maximum curvature $\kappa_{\rm max} = u_{\psi, {\rm max}}/v$.
The unsigned curvature of the trajectory of system \eqref{eq:dynamics} is
\begin{equation}
    \label{eq:curvature_2d}
    \kappa(t) = \frac{|\dot{q}_x\ddot{q}_y - \dot{q}_y\ddot{q}_x|}{(\dot{q}_x^2 + \dot{q}_y^2)^{2/3}}\;.
\end{equation}
The Dubins model can approximate fixed-wing aircraft 
\cite{Skoglar2012, Cook2014, Tyagi2021, Anand2023, Lin2022, Srinivasu2022}
and multirotors \cite{Franzetti2020}. When used for multirotors, Dubins paths ensure smoother changes in velocity making them appropriate for high-speed path planning.

\subsection{Environment Model}
\label{sec:environment}
\begin{figure}[h]
    \centering
    \includegraphics[width=1.3in]{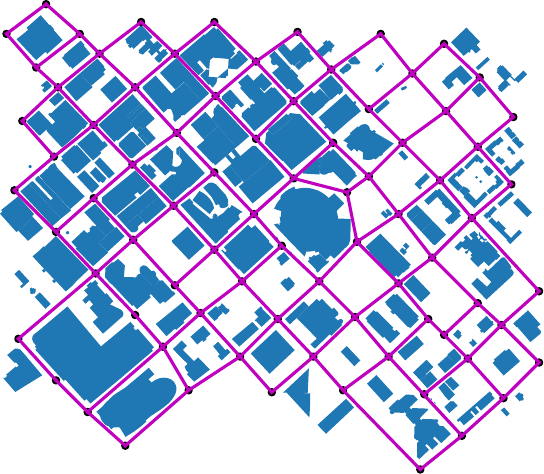} ~
    \includegraphics[width=1.3in]{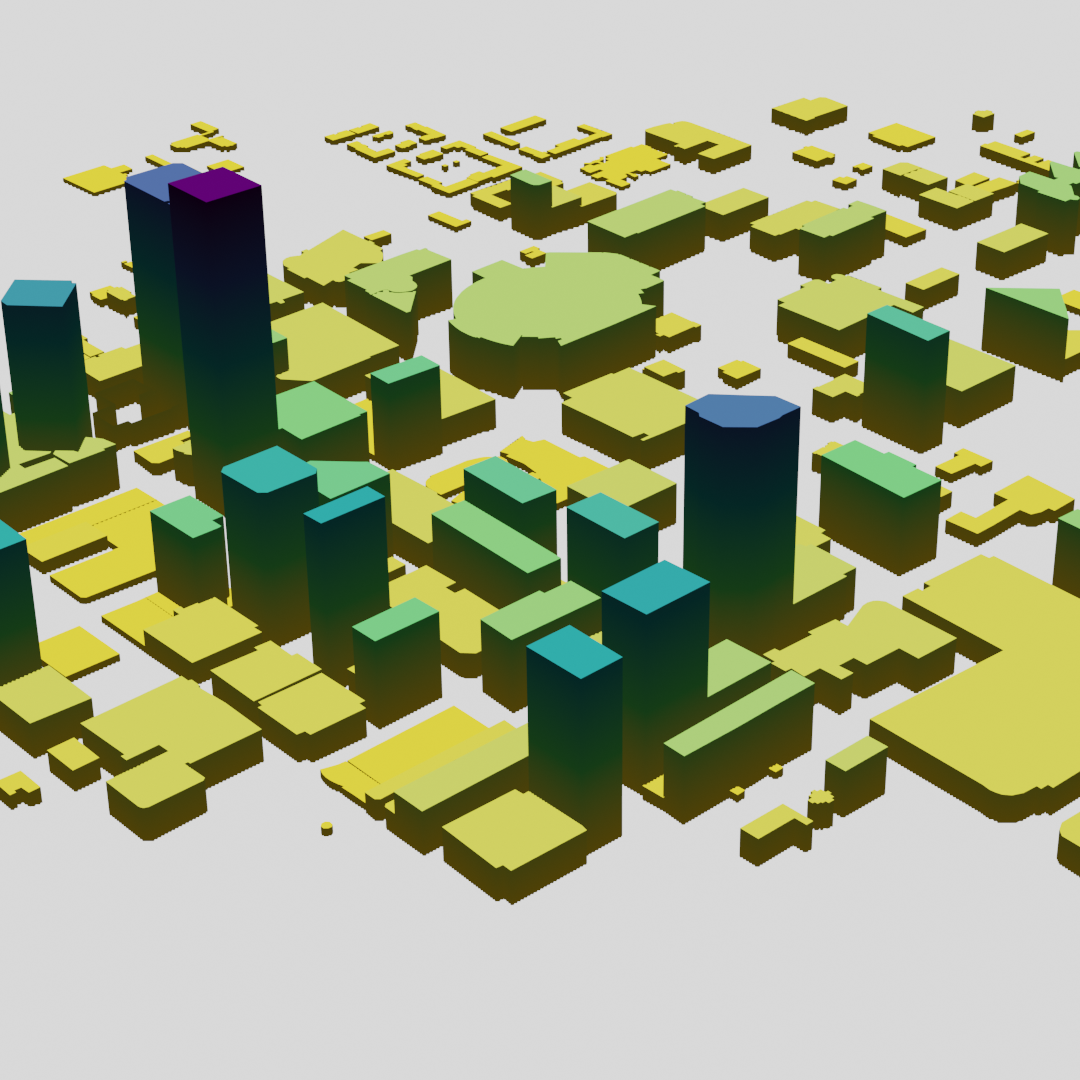}
    \caption{An example of the environment where a UAV tracks a point of interest. }
    \label{fig:environment}
\end{figure}

The UAV operates in an urban environment that consists of a ground plane with a road network and a collection of 2.5-dimensional objects representing buildings or other structures, $B$.
The $i$th object, in the set $B$ of $N_B$ objects, is an extruded polygon $B_i = \{(x,y,z) \in \mathbb{R}^3~|~(x,y) \in A_i$ and $z \in [0, h_i]\}$, where $A_i$ is a simple two-dimensional polygon, and $h_i$ is the height of the obstacle.
The boundary of polygon $A_i$ is denoted $\partial A_i$, an ordered set of points with a positive signed area.
The set of points on the interior of $A_i$ is denoted ${\rm int}(A_i)$. 
The polygonal areas of each object do not intersect, ${\rm int}(A_i) \cap {\rm int}(A_j) = \emptyset$ for all $i\neq j$ with $i,j \in \rangeset{N_B-1}$ where $\emptyset$ is the empty set.
The polygons rest on the ground plane, $z=0$ (changes in elevation are not considered). An example 2.5D environment is shown in Fig.~\ref{fig:environment}, where the ground plane is gray and the buildings are colored yellow to purple based on height.

The road network is a graph $G$ with a set of nodes $N$ and bidirectional edges $E$. Each node is assigned a position along the road and edges are assigned between adjacent nodes when connected by a road. This allows the POI to move between nodes on the graph to approximate traveling along a road. The $i$th node has a corresponding point $\bm{p}_i \in \mathbb{R}^2$ located on the ground plane in the 2.5D urban environment. Each node is unique $\bm{p}_i\neq \bm{p}_j$. The edges, defined by pairs of nodes $(i, j)$, cannot intersect with any obstacle, $l(\bm{p}_i, \bm{p}_j) \cap A_k = \emptyset$ for all $(i, j) \in E$ and $k \in \rangeset{N_B - 1}$, where $l(\bm{p}_i, \bm{p}_j) = \{\bm{p}_i + \tau(\bm{p}_j - \bm{p}_i)~{\rm for}~\tau \in [0,1]\}$ is the line segment between nodes. A road network where nodes are only defined at intersections of two roads is shown in Fig.~\ref{fig:environment}, where the footprints of the buildings are blue and the road network is magenta.
To avoid collisions, the UAV's flight path is constrained to a feasible airspace above obstacles. The UAV's feasible airspace $F = \mathbb{R}^2 \times (h_{\rm building}, h_{\rm feasible})$ is the region of space between the height of the tallest building, $h_{\rm building}$, and the maximum feasible altitude, $h_{\rm feasible}$. Accordingly, $h_{\rm UAV} \in (h_{\rm building}, h_{\rm feasible})$.

\subsection{Point of Interest Model}
\label{sec:POI_model}
The POI has a known time-varying trajectory $\bm{g}(t) = [g_x(t), g_y(t)]^{\rm T}$ that is constrained to the road network $G$, as shown in Fig.~\ref{fig:environment}, and travels along the edges (roads) at a fixed speed $||\dot{\bm{g}}||=v_{g}$. An edge $(i, j)$ between node $i$ and node $j$ is a straight line path $l(\bm{p}_i, \bm{p}_j)$. The POI's trajectory along this path is
\begin{equation}
    \bm{g}(t) = \frac{\bm{p}_j - \bm{p}_i}{||\bm{p}_j - \bm{p}_i||}v_{g}(t - t_i) + \bm{p}_i\; \notag
\end{equation}
for $t_i \in [t_i, t_j]$ where $t_i$ is the time the POI is at the $i$th node. The POI will travel along this edge until it reaches the $j$th node. 
When the POI reaches the $j$th node, the POI transitions to a new edge $(j, k) \in E$ such that $k \neq j$.
\subsection{Measurement Model}
\label{sec:measurement}
The UAV is equipped with an imaging sensor (e.g., RGB, infrared, or hyperspectral camera) mounted on a three-axis pan-tilt-roll gimbal that can regulate the camera to point at the ground plane. The camera can measure the POI when it is within sensing range and not obscured by obstacles. The region of space where the camera may image a point $\bm{g}$ is called a visibility volume (VV). When the POI is moving, the VV is time-varying. To approximate the time-varying VV, a discrete set of VVs is created where each VV corresponds to a different time. 
The line segment between the UAV at position ${\bm \rho}$ and the POI at position $\bm{g}$ is
$
    L(\tau; {\bm g}, {\bm \rho})= \tau [{\bm g}^{\rm T}~0]^{\rm T}
    + (1 - \tau){\bm \rho}
    \;,
    $
where $\tau \in [0, 1]$ is a parameter. The VV is a set of points ${\bm \rho} = [\rho_x, \rho_y, \rho_z]^{\rm T} \in F$ that have a direct line-of-sight to each target (i.e., not obscured by buildings) and satisfy other sensing constraints.  
For a gimbaled camera, the VV for a target located at ${\bm g}$ is the subset of the feasible airspace $F$ lies within direct line-of-sight to the target, and within a maximum range $d_{\rm max}$ relative to the target:
\begin{align}
        \alpha({\bm g}, F, B, d_{\rm max}) & =
        \{ {\bm \rho} \in F ~\text{such that}~
        ||{\bm \rho} - {\bm g} || \leq d_{\rm max} \notag\\&\text{and}~
         L(\tau; {\bm g}, {\bm \rho}) \cap B_i = \emptyset~\text{for all}~\tau \in[0,1]\notag\\&\text{and}~i \in \rangeset{N_B-1} \}\;.
    \label{eq:visibility_volumes}
\end{align}
For brevity, VVs \eqref{eq:visibility_volumes} are denoted $\alpha({\bm g})$. 
The maximum range $d_{\rm max}$ is determined by the minimum pixel requirements for imaging a target of a known size for a desired resolution.
VVs are computed using \cite{Hague2023}, resulting in a triangular mesh.

\subsection{Problem Statement: Reference Orbit Design}
\label{sec:problem_state}
An orbit may serve as a reference path for a UAV continuously encircling the POI. The reference orbit should be inside the POI's VV as the POI travels through the environment over a time interval $[t_0, t_f]$ where $t_f > t_0$. By following the VO, inside the VV, the UAV maintains a view of the target at all times. Let $O(\theta, t)~:~\mathbb{S} \times \mathbb{R} \to \mathbb{R}^2$ denote a time-varying orbit that is parameterized by the angle $\theta$ about the moving point $\bm{g}(t) = [g_x(t)~g_y(t)]^\T$ and the time $t$. For each point in time $t$, the orbit function $O(\theta,t)$ produces a closed non-self-intersecting curve in the $xy$ plane when sweeping out the angle $\theta \in \mathbb{S}$. The orbit is centered on ${\bm g}(t)$ and is located at a constant altitude $z = h_{\rm UAV}$.
The curvature of the orbit in the inertial frame, denoted  $\kappa(O(\theta, t), \theta, t)$ at any angle $\theta$ and time $t$, must remain less than the maximum curvature $\kappa_{\rm max}$ to be feasible for the system \eqref{eq:dynamics}. The orbit must also remain in the POI's VV, $\alpha(\bm{g}(t))$.
While many types of orbit geometries are feasible, this work considers circular orbits. The time-varying orbit is defined as
\begin{equation}
O(\theta,t) = 
\begin{bmatrix}
R(t)\cos \theta + g_x(t) \\
R(t)\sin \theta + g_y(t)
\end{bmatrix} =\begin{bmatrix}
    O_x(\theta, t)\\
    O_y(\theta, t)
\end{bmatrix}\;
\label{eq:orbit_radius_parametrized}
\end{equation}
for $\theta \in \mathbb{S}$ where $R(t)$ is a time-varying radius. 
The standoff distance (i.e., the radius of the orbit) is maximized to reduce the curvature and observe the sides of the POI. The optimization problem is to specify $R(t)$ so that the orbit remains feasible for the Dubins vehicle model \eqref{eq:dynamics}, guarantees the visibility of the POI, and maximizes the standoff distance:
\begin{equation}
    {\rm{maximize}} \quad 
    R(t)
    \label{eq:optimization_prob}
\end{equation}
for all $t \in [t_0, t_f]$
subject to
\begin{align}
    [O_x(\theta, t), O_y(\theta, t), h_{\rm UAV}]^\T & \subset \alpha(\bm{g}(t)) \label{eq:in_visibility}\\
    \kappa(O(\theta, t), \theta, t) &\leq \kappa_{\rm max}~{\rm for~all}~\theta \in \mathbb{S} \label{eq:curvature_constraint_optimization}\\
    \bm{g}(t_0) &= \bm{p}_i~{\rm for}~ i \in N\label{eq:POI_start}\\
    \bm{g}(t_f) &= \bm{p}_j~{\rm for}~j \in N\label{eq:POI_end} \;,
\end{align}
where $\bm{g}(t)$ is a known continuous path through the environment, $\bm{p}_i$ is the location of the $i$th (initial) node on the road network, and $\bm{p}_j$ is the final  node.
The first constraint \eqref{eq:in_visibility} ensures the orbit is contained inside the VV. The curvature constraint \eqref{eq:curvature_constraint_optimization} ensures that the UAV's turn rate constraint \eqref{eq:dynamics_constraint} is satisfied on the orbit. The final constraints \eqref{eq:POI_start} and \eqref{eq:POI_end} specify the POI start and end positions on the road network $G$.

\subsection{Problem Statement: Orbit Controller Design}
\label{sec:form_control}
The control objective is to design a controller $u_\psi(t, \bm{q})$ that steers the system \eqref{eq:dynamics} onto the orbit $O(\theta, t)$, $\lim_{t \rightarrow \infty} ~ [q_x(t), q_y(t)]^{\rm T} - O_{\rm min}(t, \bm{q}(t)) \rightarrow \bm{0}$, where $O_{\rm min}(t, \bm{q}(t)) \in \mathbb{R}^2$ is the closest point on the orbit to the configuration $\bm{q}(t)$, and the point $[q_x(t), q_y(t)]^{\rm T}$ is the inertial position component of the configuration $\bm{q}(t)$.
In the case where the UAV is in the middle of the circular orbit, multiple points are equidistant from the internal position $[q_x(t), q_y(t)]^{\rm T}$ and the UAV is attracted to the point on the orbit in the direction of the UAV's heading.

\section{Moving View Polygons}
\label{sec:polygons}

\begin{figure}[h]
    \centering
    \includegraphics[width=.25\textwidth]{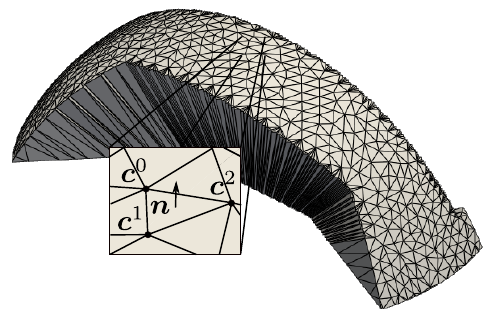}
    \caption{A VV approximated as a triangular mesh, each triangular face element has three nodes ordered so that the cross-product of the nodes points out of the mesh.}
    \label{fig:mesh}
\end{figure}

As the POI moves over the road network its VV changes continuously. The moving VV is represented by a series of static VVs at different points $\mathcal{P} = \{ {\bm p}_0, \ldots, {\bm p}_{m -1}\}$ along the POI's known path at the corresponding times $t_{0:m-1} = \{ t_0, \ldots, t_{m -1} \}$. The VVs at these points are calculated and approximated with the gimballed model in \cite{Hague2023}. The resulting meshes are a set of nodes and triangular faces, denoted $\hat{\alpha}(\bm{p}_i)$ (see Fig.~\ref{fig:mesh}) for each ${\bm p}_i \in \mathcal{P}$.
To solve the orbit design problem \eqref{eq:optimization_prob}, a piecewise linear time-varying radius $R(t)$ is considered. For each point ${\bm p}_i \in \mathcal{P}$ a VP is computed at the altitude $h_{\rm UAV}$ and a corresponding radius $R_i$ is determined as the largest circle centered on ${\bm p}_i$ inscribed in the VP.

First, this section shows the conditions under which a given orbit $R(t)$ \eqref{eq:orbit_radius_parametrized} is feasible for the Dubins vehicle \eqref{eq:dynamics}. Then, to better optimize the orbit $R(t)$, an adaptive discretization approach determines the locations $\mathcal{P}$ at which to compute the VVs, reducing changes in VV geometry. The VVs are then sliced along a constant altitude plane $z=h_{\rm UAV}$ with \cite{Jiantao2004} to produce the VPs.
Finally, an algorithm is presented to create the final time-varying orbit for the POI's path while considering feasibility constraints.
From this section forward, function parameters will be suppressed for brevity.

\subsection{Feasibility of Reference Orbits}
\label{sec:feasibility}

The minimum turning radius of the Dubins vehicle in an inertial frame is $r_{\rm min}$. However, even if $R\geq r_{\rm min}$ the orbit may not be feasible for the system \eqref{eq:dynamics} since the inertial path curvature of a translating circle with radius $R$ is different from a non-translating circle with radius $R$. This section derives conditions for the feasibility of an orbit, defined by a piecewise linear time-varying radius $R$, for the system \eqref{eq:dynamics}. 
\begin{defin}
    \label{def:poi}
    Let $\bm{g} = \dot{\bm{g}}(t-t_i) + \bm{g}_i$ be a time-parametrized straight-line trajectory on the ground plane starting at $\bm{g}_i \in \mathbb{R}^2$ at time $t_i$ with constant velocity $\dot{\bm{g}} \in \mathbb{R}^2$ and speed $v_g = ||\dot{\bm{g}}||$. The vector $\dot{\bm{g}}$ is expressed in rectangular coordinates or in magnitude-heading form as
    \begin{equation}
        \dot{\bm{g}}=\begin{bmatrix}
            \dot{g}_x \\
            \dot{g}_y
        \end{bmatrix} =
        v_g\begin{bmatrix}
            \cos\gamma\\
            \sin\gamma
        \end{bmatrix}\;,
        \label{eq:poi_velocity}
    \end{equation}
    where $\gamma$ is the heading of the POI.
\end{defin}
\begin{defin}
    \label{def:radius}
    Let 
    \begin{equation}
        R = R_i\left(\frac{t_j - t}{t_j - t_i}\right) + R_j\left(\frac{t -t_i}{t_j - t_i}\right) \label{eq:linear_orbit}
    \end{equation}
     be a time-varying radius valid over the time interval $t \in [t_i, t_j]$ where $R_i, R_j > 0$ are radii corresponding to the start and end of the time interval.
\end{defin}
\begin{defin}
    \label{def:polar}
    Let $\bm{e}_r = [\cos \theta, \sin \theta]^\T$ and $\bm{e}_\theta = [-\sin \theta, \cos \theta]^\T$ be orthogonal polar-coordinate unit vectors measured from the point $\bm{g}$ where $\theta = \arctan \left(\frac{q_y - g_y}{q_x - g_x}\right)$ is the angle from the moving POI to the point $(q_x, q_y)$ measured counter-clockwise from a vector parallel to the $x$-axis.
\end{defin}
\begin{lem}
\label{lem:UAV_polar_speed}
If the system \eqref{eq:dynamics} travels along the locus of orbits $O$ described by \eqref{eq:orbit_radius_parametrized} and \eqref{eq:linear_orbit} then the polar angle rate-of-change is
\begin{equation}
    \dot \theta
    = \frac{1}{R}\left(-\dot{\bm{g}}\cdot\bm{e}_\theta \pm \sqrt{v^2 - (\dot{\bm{g}}\cdot\bm{e}_r + \dot{R})^2}\right)\;,
    \label{eq:ang_vel}
\end{equation}
where the ``+'' represents counter-clockwise (CCW) motion about $\bm{g}$ and the ``-'' represents clockwise (CW) motion.
\begin{proof}
Suppose that the inertial coordinates of the system \eqref{eq:dynamics} begin on the orbit $O$ at an angle $\theta$ about $\bm{g}$. The inertial location is described by \eqref{eq:orbit_radius_parametrized} and \eqref{eq:linear_orbit} for time $t \in [t_i, t_j]$ and can be expressed as
\begin{equation}
    \bm{x}_\mathcal{I} = \bm{g} + R \bm{e}_r\;.
    \label{eq:xI}
\end{equation}
The time derivative of \eqref{eq:xI} considering \eqref{eq:linear_orbit} is
\newcommand{\rprime}{R'_i\left(\frac{t_j - t}{t_j - t_i}\right) + R'_j\left(\frac{t - t_i}{t_j - t_i}\right)}
\newcommand{\rfunc}{R_i\left(\frac{t_j - t}{t_j - t_i}\right) + R_j\left(\frac{t - t_i}{t_j - t_i}\right)}
\newcommand{\rdot}{\frac{1}{t_j - t_i}\left(R_j - R_i\right)}
\begin{equation}
\dot{\bm{x}}_{\mathcal{I}} = \dot{\bm{g}}
    + \dot R \bm{e}_r + R \dot{\theta}{\bm e}_\theta \;, \label{eq:velocity_generic}
\end{equation}
where the rate of change of the radius is
\begin{equation}
    \dot{R} = \rdot\;.
    \label{eq:dotR_termsRiRj}
\end{equation}
By using Definition~\ref{def:polar}, the velocity \eqref{eq:velocity_generic} is expanded to
\begin{equation}
    \dot{\bm{x}}_\mathcal{I} = \begin{bmatrix}
        \dot{g}_x\\
        \dot{g}_y
    \end{bmatrix} +
    \dot{R}
    \begin{bmatrix}
        \cos \theta \\
        \sin \theta 
    \end{bmatrix} +
    R \dot{\theta}
    \begin{bmatrix}
        -\sin \theta  \\
        \cos \theta 
    \end{bmatrix}\;. \notag
\end{equation}
Since the system \eqref{eq:dynamics} has a constant speed, it is required that 
\begin{align}
        ||\dot{\bm{x}}_\mathcal{I}||^2 = v^2  & = \left(\dot{g}_x + \dot{R}\cos \theta  - R\dot{\theta}\sin \theta \right)^2  + \left(\dot{g}_y + \dot{R}\sin \theta  + R\dot{\theta}\cos{\theta}\right)^2 \;
        \label{eq:xIdotsq}
\end{align}
which can be re-written as a  quadratic equation in $\dot \theta$, 
\begin{align}
    0 & = (v_{g}^2 - v^2 + 2\dot{R}\dot{\bm{g}}\cdot \bm{e}_r + \dot{R}^2) + \left( 2R\dot{\bm{g}}\cdot \bm{e}_\theta\right)\dot{\theta} +  R^2 \dot{\theta}^2
    \label{eq:quadratic_theta}\;,
\end{align}
where $(\cdot)$ represents the standard dot product of two vectors.
Solving \eqref{eq:quadratic_theta} for $\dot{\theta}$ leads
to the polar angle rate-of-change,
\begin{equation}
    \dot \theta = \frac{1}{R}\left(-\dot{\bm{g}}\cdot\bm{e}_\theta \pm \sqrt{(\dot{\bm{g}}\cdot\bm{e}_r)^2 - \left(v_{g}^2 - v^2 + 2\dot{R}\dot{\bm{g}}\cdot \bm{e}_r + \dot{R}^2\right)}\right)\;.
    \label{eq:ang_vel_a}
\end{equation}
By rearranging the term under the square root in \eqref{eq:ang_vel_a}, equation \eqref{eq:ang_vel} is found.
\end{proof}
\end{lem}
\begin{remark}
    \label{rem:theta_t}
    Equation \eqref{eq:ang_vel} is a time-varying differential equation valid over the interval $t\in[t_i, t_j]$ for any initial condition $\theta(t_i)=\theta_i$ and direction (CW or CCW) that defines a unique constant-speed trajectory remaining on the locus of orbits defined by $R$ and $\bm{g}$. The resulting path implicitly determines the vehicle heading on this orbit.
\end{remark}
\begin{defin}
    \label{def:remain}
    Let ``remain on the orbit $O$'' be the condition wherein the angular velocity $\dot{\theta}$ about $\bm{g}$ is \eqref{eq:ang_vel} and greater or equal to zero for CCW circulation and less than or equal to zero for CW circulation for all $\theta \in \mathbb{S}$.
\end{defin}

\begin{lem}
\label{lem:curvature_polar}
Given a trajectory $\bm{x}_\mathcal{I}\in \mathbb{R}^2$ with velocity $\dot{\bm{x}}_\mathcal{I} \in \mathbb{R}^2$, constant speed $v=||\dot{\bm{x}}_\mathcal{I}||$, and acceleration $\ddot{\bm{x}}_\mathcal{I} \in \mathbb{R}^2$ expressed in a polar frame according to Definition~\ref{def:polar}, the signed curvature of the trajectory $\bm{x}_\mathcal{I}$ is 
\begin{equation}
    \kappa_s = \frac{1}{v}\left(-\frac{\frac{\rm d}{{\rm d}t}\left(\dot{\bm{x}}_\mathcal{I}\cdot\bm{e}_r\right)}{\dot{\bm{x}}_\mathcal{I}\cdot\bm{e}_\theta} + \dot{\theta}\right)\;.
    \label{eq:curvature_polar}
\end{equation}
    \begin{proof}
        The velocity $\dot{\bm{x}}_\mathcal{I}$ can be resolved as
\begin{equation}
    \dot{\bm{x}}_\mathcal{I} = a \bm{e}_r + b\bm{e}_\theta\;,
    \label{eq:ab_sub}
\end{equation}
where $a$ and $b$ are scalar components. The derivative with respect to time $\frac{\rm d}{{\rm d}t}\dot{\bm{x}}_\mathcal{I}$ is
\begin{align}
    \ddot{\bm{x}}_\mathcal{I}
    & = (\dot{a} -b\dot{\theta})\bm{e}_r + (\dot{b} + a\dot{\theta})\bm{e}_\theta \;,  
    \label{eq:ab_sub_dot}
\end{align}
where the subscript notation $\ddot{x}_{\mathcal{I},r}$ and $\ddot{x}_{\mathcal{I},\theta}$ indicate the polar components of the acceleration. The scalars $\ddot{x}_{\mathcal{I},x}$ and $\ddot{x}_{\mathcal{I},y}$ are the rectangular components of the acceleration and are related to the polar components by 
\begin{equation}
    \begin{bmatrix}
    \ddot{x}_{\mathcal{I},x} \\
    \ddot{x}_{\mathcal{I},y}
    \end{bmatrix}
    =
        \begin{bmatrix}
    \cos \theta & -\sin \theta \\
    \sin \theta & \cos \theta
    \end{bmatrix}
    \begin{bmatrix}
    \ddot{x}_{\mathcal{I},r} \\
    \ddot{x}_{\mathcal{I},\theta}
    \end{bmatrix}  \;.  
\end{equation}
There is a similar relation for $\dot{x}_{\mathcal{I},r}$ and $\dot{x}_{\mathcal{I},\theta}$ with $\dot{x}_{\mathcal{I},x}$ and $\dot{x}_{\mathcal{I},y}$, resulting in an expression for signed curvature,
\begin{align}
     \kappa_s %
    & = \frac{\ddot{x}_{\mathcal{I}, \theta} \dot{x}_{\mathcal{I}, r} - \ddot{x}_{\mathcal{I}, r}\dot{x}_{\mathcal{I}, \theta}}{\left(\dot{x}_{\mathcal{I}, r}^2 + \dot{x}_{\mathcal{I}, \theta}^2\right)^{3/2}}\;. \label{eq:curvature_polar_varying_vel}
\end{align}
The velocity term in the denominator is invariant under coordinate system transformations. By using the equation for the signed curvature \eqref{eq:curvature_polar_varying_vel} and considering the formulations \eqref{eq:ab_sub} and \eqref{eq:ab_sub_dot} the curvature can be expressed as
\begin{equation}
    \kappa_s  = \frac{1}{v^3}\left(a\dot{b} - \dot{a}b + \dot{\theta}(a^2 + b^2)\right)\;.
    \label{eq:curvab}
\end{equation}
The constraint $||\dot{\bm{x}}_\mathcal{I}|| = v = \sqrt{a^2 + b^2}$ implies that \eqref{eq:curvab} is
\begin{equation}
    \kappa_s = \frac{1}{v^3}\left(a\dot{b} - \dot{a}b\right) + \frac{\dot{\theta}}{v}\;.
    \label{eq:dot_psi_d_partial_simp}
\end{equation}
Since $v^2 = a^2 + b^2$ is a constant, then $\frac{\rm d}{{\rm d}t}v^2 = \frac{\rm d}{{\rm d}t}\left(a^2 + b^2\right) =  2\dot{a}b + 2\dot{b}b = 0$ and it follows that 
\begin{align}
    \dot{b} &= -\frac{\dot{a}a}{b}\;.
    \label{eq:controller_deriviative_theta_sub}
\end{align}
Substituting \eqref{eq:controller_deriviative_theta_sub} into \eqref{eq:dot_psi_d_partial_simp},
\begin{align}
    \kappa_s & = \frac{1}{v^3}\left(-\frac{a^2\dot{a}}{b} - b\dot{a}\right) + \frac{\dot{\theta}}{v}\notag\\
    & = \frac{1}{v}\left(-\frac{\dot{a}}{b} + \dot{\theta}\right) \label{eq:ab_curvature}\;,
\end{align}
a simplified form of $\kappa_s$ is found. 
The value of $a$ is the component of $\dot{\bm{x}}_\mathcal{I}$ in the $\bm{e}_r$ direction from \eqref{eq:ab_sub}, thus
\begin{equation}
    \dot{a} = \frac{\rm d}{{\rm d}t}(\dot{\bm{x}}_\mathcal{I}\cdot \bm{e}_r)\;.
    \label{eq:a_value}
\end{equation}
Similarly,
\begin{equation}
    b = \dot{\bm{x}}_\mathcal{I}\cdot\bm{e}_\theta\;
    \label{eq:b_value}
\end{equation}
is the component of $\dot{\bm{x}}_\mathcal{I}$ in the $\bm{e}_\theta$ direction. Substituting \eqref{eq:a_value} and \eqref{eq:b_value} into \eqref{eq:ab_curvature} gives \eqref{eq:curvature_polar}.
    \end{proof}
\end{lem}
\begin{lem}
    \label{lem:curavture_constant}
If the system \eqref{eq:dynamics} remains on the orbit $O$ then the unsigned curvature of the path \eqref{eq:curvature_2d} is
\begin{equation}
    \label{eq:orbit_curvature_constraint}
    \kappa = \frac{R\dot{\theta}^2}{v\sqrt{v^2 - (\dot{\bm{g}}\cdot{\bm{e}_r + \dot{R})^2}}} \leq \kappa_{\rm max}\;.
\end{equation}
\begin{proof}
Lemma~\ref{lem:curvature_polar} applies to system \eqref{eq:dynamics}, because system \eqref{eq:dynamics} has constant velocity; Therefore, the signed curvature of a trajectory is \eqref{eq:orbit_curvature_constraint} and the unsigned curvature is
\begin{equation}
    \kappa = \frac{1}{v}\left|-\frac{\frac{\rm d}{{\rm d}t}\left(\dot{\bm{x}}_\mathcal{I}\cdot\bm{e}_r\right)}{\dot{\bm{x}}_\mathcal{I}\cdot\bm{e}_\theta} + \dot{\theta}\right|\label{eq:curvature_lemma}\;.
\end{equation}
By expressing the velocity $\dot{\bm{x}}_\mathcal{I}$ from \eqref{eq:velocity_generic} as 
\begin{equation}
    \dot{\bm{x}}_\mathcal{I}=(\dot{\bm{g}}\cdot\bm{e}_r + \dot{R})\bm{e}_r + (\dot{\bm{g}}\cdot\bm{e}_\theta + R\dot{\theta})\bm{e}_\theta\;,
    \label{eq:xIdot_grouped}
\end{equation}
the derivative of $\dot {\bm x}_\mathcal{I}\cdot {\bm e}_r$ is  
\begin{align}
    \frac{\rm d}{{\rm d}t}\left(\dot{\bm{x}}_\mathcal{I}\cdot\bm{e}_r\right) & = \ddot{\bm{g}}\cdot\bm{e}_r 
 + \dot{\theta}\dot{\bm{g}}\cdot\bm{e}_\theta\ + \ddot{R}\notag\\ &=\dot{\theta}\dot{\bm{g}}\cdot\bm{e}_\theta\;,
    \label{eq:acceleration}
\end{align}
since $\ddot{R} = 0$ for $t \in [t_i, t_j]$ under the assumption of Definition~\ref{def:radius} and $\ddot{\bm{g}} = 0$ under the assumption of Definition~\ref{def:poi}.
By substituting %
\eqref{eq:xIdot_grouped} 
and \eqref{eq:acceleration}  into \eqref{eq:curvature_lemma} the unsigned curvature is
\begin{align}
    \kappa&=\frac{1}{v}\left|-\frac{\dot{\theta}\dot{\bm{g}}\cdot\bm{e}_\theta}{\dot{\bm{g}}\cdot\bm{e}_\theta + R\dot{\theta}} + \dot{\theta}\right|\label{eq:curvature_signed_thetadot}\\
    &=\frac{1}{v}\left|\frac{\dot{\theta}\left(-\dot{\bm{g}}\cdot\bm{e}_\theta + \dot{\bm{g}}\cdot\bm{e}_\theta + R\dot{\theta}\right)}{\dot{\bm{g}}\cdot\bm{e}_\theta + R\dot{\theta}}\right|\;.
    \label{eq:curvature_unsimplified}
\end{align}
Equation \eqref{eq:xIdot_grouped} with $v = ||{\bm x}_\mathcal{I}||$ is re-written as   $R\dot{\theta} = -\dot{\bm{g}}\cdot\bm{e}_\theta \pm \sqrt{v^2 - (\dot{\bm{g}}\cdot\bm{e}_r + \dot{R})^2}$. This fact,  along  with \eqref{eq:ang_vel} that holds along the orbit, allows the denominator in \eqref{eq:curvature_unsimplified} to be expanded and simplified:
\begin{align}
    \kappa&=\frac{1}{v}\left|\frac{R\dot{\theta}^2}{\dot{\bm{g}}\cdot\bm{e}_\theta - \dot{\bm{g}}\cdot\bm{e}_\theta \pm \sqrt{v^2 - (\dot{\bm{g}}\cdot\bm{e}_r + \dot{R})^2}}\right|\notag\\
    &=\frac{R\dot{\theta}^2}{v\sqrt{v^2 - (\dot{\bm{g}}\cdot{\bm{e}_r + \dot{R})^2}}}\;.
    \label{eq:orbit_curvature}
\end{align}
For the constraints \eqref{eq:dynamics_constraint} to be satisfied the  unsigned curvature must be bounded, resulting in \eqref{eq:orbit_curvature_constraint}.
\end{proof}
\end{lem}

\begin{remark}
    \label{rem:dubins_traj}
    If a trajectory $\bm{x}_\mathcal{I}$ has constant speed $v$ and bounded unsigned curvature $\kappa \leq \kappa_{\rm max}$ then the system \eqref{eq:dynamics} can follow the trajectory. The Dubins model can be written as \eqref{eq:dynamics} with $u_\psi= v\kappa_s$
    Therefore, a trajectory is uniquely defined by an initial condition,
    $        \bm{q} =[
            q_{x, 0},
            q_{y, 0},
            q_{\psi, 0}]^{\rm T}$,
    and a curvature control  $|\kappa_s| < \kappa_{\rm max}$.
\end{remark}
\begin{lem}
    \label{lem:remain_orbit}
    Suppose the system \eqref{eq:dynamics} starts on the orbit with initial conditions:
    \begin{align}
        \label{eq:init_speed}
        \begin{bmatrix}
            \dot{q}_{x}(t_0)\\
            \dot{q}_{y}(t_0)
        \end{bmatrix} & = \dot{\bm{g}}(t_0) + \dot R(t_0)\bm{e}_r(t_0) + R(t_0)\dot \theta(t_0)\bm{e}_\theta(t_0) \;,\\
        q_{\psi}(t_0) &= \arctan\left(\frac{\dot{q}_{y}(t_0)}{\dot{q}_{x}(t_0)}\right) \;, \label{eq:init_heading} \\
        \bm{q}(t_0) & =\begin{bmatrix}g_x(t_0)+R(t_0)\cos\theta_0\\ g_y(t_0) + R(t_0)\sin\theta_0\\ q_{\psi}(t_0)\end{bmatrix}\label{eq:init_config}\;,
    \end{align}
    where the choice of circulation direction CW or CCW is determined by the sign choice of the $\pm$ term for  $\dot \theta(t_0)$ in  \eqref{eq:ang_vel}. Then, if the system $\eqref{eq:dynamics}$ uses the control $u_\psi=v\kappa_s$ from \eqref{eq:orbit_curvature_constraint},
    the system remains on the orbit.

    \begin{proof}
        Let the distance from the orbit be
        $
            \tilde{r} = R - \sqrt{(g_x - q_x)^2 + (g_y -q_y)^2}
        $.    
        The first derivative of the distance is
        \begin{equation}
            \dot{\tilde{r}} = \dot R - \frac{(\dot{g}_x - \dot q_x)(g_x -q_x) + (\dot{g}_y - \dot q_y)(q_y - g_y)}{\sqrt{(g_x - q_x)^2 + (g_y -q_y)^2}}\;.\notag
        \end{equation}
        By substituting $\bm{e}_r$ and $\bm{e}_\theta$,
        \begin{equation}
            \dot{\tilde{r}} = \dot R - \dot{\bm{g}}\cdot \bm{e}_r+\begin{bmatrix}
                \dot{q}_x\\
                \dot{q}_y
            \end{bmatrix}\cdot \bm{e}_r\;.
            \label{eq:dot_delta2}
        \end{equation}
        Inserting $\dot q_x$ and $\dot q_y$ from   \eqref{eq:dynamics},
        \begin{equation}
            \dot{\tilde{r}} = \dot R -\dot{\bm{g}}\cdot \bm{e}_r+v\begin{bmatrix}
                \cos q_\psi\\
                \sin q_\psi
            \end{bmatrix}\cdot \bm{e}_r\;.
            \label{eq:dot_delta}
        \end{equation}
        The second derivative is
        \begin{equation}
            \ddot{\tilde{r}} = -\dot \theta \dot {\bm g} \cdot \bm{e}_\theta +vu_\psi\begin{bmatrix}
                -\sin q_\psi\\
                \cos q_\psi
            \end{bmatrix}\cdot \bm{e}_r + v\dot\theta\begin{bmatrix}
                \cos q_\psi\\
                \sin q_\psi
            \end{bmatrix}\cdot \bm{e}_\theta\;.\notag
        \end{equation}
        By using the relation between curvature and control, $\kappa_s v = u_\psi$, 
        
        \begin{equation}
            \ddot{\tilde{r}} = -\dot \theta \dot {\bm g} \cdot \bm{e}_\theta -v^2\kappa_s(\sin q_\psi \cos\theta - \cos q_\psi \sin\theta)+ v\dot\theta(\sin q_\psi\cos\theta -\cos q_\psi\sin\theta)\;.
            \label{eq:delta_ddot2}
        \end{equation}
        Simplifying the trigonometric functions,
        \begin{equation}
            \ddot{\tilde{r}} = -\dot \theta \dot {\bm g} \cdot \bm{e}_\theta \
        + (v\dot\theta - v^2\kappa_s)\sin(q_\psi - \theta)\;.
            \label{eq:delta_ddot}
        \end{equation}
        The signed variant of the curvature \eqref{eq:orbit_curvature_constraint} can be written as 
        \begin{equation}
            \kappa_s=\frac{1}{v}\left(\pm\frac{-\dot{\theta}\dot{\bm{g}}\cdot\bm{e}_\theta}{\sqrt{v^2 - \left(\dot{\bm{g}}\cdot\bm{e}_\theta + \dot R\right)^2}} + \dot{\theta}\right)\;.
            \label{eq:singed_curvature_theta}
        \end{equation}
        Substituting \eqref{eq:singed_curvature_theta} into \eqref{eq:delta_ddot},
        \begin{equation}
            \ddot{\tilde{r}} = \pm\dot\theta\dot{\bm{g}}\cdot\bm{e}_\theta v\left(v^2 - \left(\dot{\bm{g}}\cdot\bm{e}_\theta + \dot R\right)^2\right)^{-1/2}\sin(q_\psi - \theta) -\dot{\theta}\dot{\bm{g}}\cdot \bm{e}_\theta\;.
            \label{eq:ddot_delta_expanded}
        \end{equation}
        By equating the inertial velocity of the orbit and UAV in the $\bm{e}_\theta$ direction,
        \begin{equation}
            \dot{\bm{x}}_\mathcal{I}\cdot \bm{e}_\theta = v\begin{bmatrix}
                \cos q_\psi\\
                \sin q_\psi
            \end{bmatrix}\cdot\bm{e}_\theta\;.
            \label{eq:theta_intertial}
        \end{equation}
        From the definition of curvature in polar coordinates \eqref{eq:curvature_polar} and the signed curvature \eqref{eq:singed_curvature_theta},
        \begin{equation}
            \dot{\bm{x}}_\mathcal{I}\cdot\bm{e}_\theta = \pm\sqrt{v^2 - \left(\dot{\bm{g}}\cdot\bm{e}_\theta + \dot R\right)^2}\;.
            \label{eq:dotxI_dot_etheta}
        \end{equation}
        Substituting \eqref{eq:theta_intertial} into \eqref{eq:dotxI_dot_etheta} and simplifying with the angle sum and difference identity, 
        \begin{equation}
            \pm\sqrt{v^2 - \left(\dot{\bm{g}}\cdot\bm{e}_\theta + \dot R\right)^2} = v\sin(q_\psi - \theta)\;.
            \label{eq:sine_term}
        \end{equation}
        Finally, substituting \eqref{eq:sine_term} into \eqref{eq:ddot_delta_expanded},
        \begin{align}
            \ddot{\tilde{r}} &=  \pm\dot\theta\dot{\bm{g}}\cdot\bm{e}_\theta v\left(v^2 - \left(\dot{\bm{g}}\cdot\bm{e}_\theta + \dot R\right)^2\right)^{-1/2} \left( \pm\frac{1}{v}\sqrt{v^2 - \left(\dot{\bm{g}}\cdot\bm{e}_\theta + \dot R\right)^2}\right) -\dot{\theta}\dot{\bm{g}}\cdot \bm{e}_\theta\notag 
            = 0\notag\;.
        \end{align}
        Since the $\pm$ terms cancel out based on a choice of direction, the direction of travel has no effect on system \eqref{eq:dynamics} remaining on the orbit.
        By assumption of that the vehicle begins on the orbit, the error $\tilde{r}(t_0)=0$. Substituting the assumed initial velocities \eqref{eq:init_speed} into \eqref{eq:dot_delta2}:        
        \begin{equation}
            \dot{\tilde{r}}(t_0)=\dot R(t_0) - \dot{\bm{g}}\cdot\bm{e}_r(t_0) + \dot{\bm{g}}\cdot\bm{e}_r(t_0) + \dot R(t_0)\bm{e}_r(t_0)\cdot\bm{e}_r(t_0) + R(t_0)\dot \theta(t_0)\bm{e}_\theta(t_0) \cdot \bm{e}_r(t_0)=0\;.\notag
        \end{equation}
        Because the radial distance to the orbit $\tilde{r}$ starts at zero, the derivative $\dot{\tilde{r}}$ starts at zero, and the second derivative $\ddot{\tilde{r}}$ is always zero, then $\tilde{r}$ is zero for all time. Thus, the system remains on the orbit $O$.
    \end{proof}
\end{lem}
\begin{theorem}
    \label{thm:orbit}
    Suppose that the system \eqref{eq:dynamics} begins on an orbit $O$ with initial conditions \eqref{eq:init_speed}--\eqref{eq:init_config}.
    The system \eqref{eq:dynamics} remains on the orbit $O$ if and only if the curvature control is given by the left-hand side (LHS) of \eqref{eq:orbit_curvature_constraint} and the inequality of \eqref{eq:orbit_curvature_constraint} is satisfied.
    \begin{proof}
        The fact that remaining on the orbit with system \eqref{eq:dynamics} implies the curvature control and bound \eqref{eq:orbit_curvature_constraint} was shown in  Lemma~\ref{lem:curavture_constant}.
        For the reverse direction, suppose the curvature control is given by the LHS of  \eqref{eq:orbit_curvature_constraint} and the bound is satisfied. By assumption, the system \eqref{eq:dynamics} begins on the orbit with the initial condition given. To show that \eqref{eq:orbit_curvature_constraint} implies remaining on the orbit we must establish that (i) \eqref{eq:orbit_curvature_constraint} is feasible for the system \eqref{eq:dynamics}, and (ii) executing the curvature control given by the LHS of \eqref{eq:orbit_curvature_constraint} with system \eqref{eq:dynamics} results in a trajectory that remains on the orbit.
        
        Part (i): The constraint \eqref{eq:orbit_curvature_constraint} implies that the curvature constraint \eqref{eq:dynamics_constraint} is satisfied.  In the proof of Lemma~\ref{lem:UAV_polar_speed} the magnitude of the velocity of the $xy$ components \eqref{eq:xIdotsq} on the trajectory is required to be $v$, thus \eqref{eq:orbit_curvature_constraint} also implies a constant speed $v$. Considering Remark~\ref{rem:dubins_traj}, the condition \eqref{eq:orbit_curvature_constraint} leads to a feasible trajectory for the system \eqref{eq:dynamics}. 
        
        Part (ii): 
        The initial condition ${\bm q}_0$ describes the polar angle $\theta_0$ and position on the orbit $O(\theta_0,t_0)$. 
        From Lemma~\ref{lem:remain_orbit}, if system \eqref{eq:dynamics} follows the curvature control \eqref{eq:orbit_curvature_constraint} with the initial condition $\bm{q}_0$, 
        then system \eqref{eq:dynamics} will remain on the orbit. 
    \end{proof}
\end{theorem}

\begin{remark}
    Theorem~\ref{thm:orbit} is satisfied when the POI travels along the straight path, introduced in Definition~\ref{def:poi}, corresponding to a road network edge, but does not hold for when the POI turns onto a new road network edge and changes the direction of the velocity $\dot{\bm{g}}$ instantaneously. Additionally, for \eqref{eq:orbit_curvature_constraint} to be satisfied, the radial rate of change $\dot{R}$ must be sufficiently small. The acceleration $\ddot{\bm{x}}_\mathcal{I}$ and curvature $\kappa$ are infinite for discontinuous $\dot{\bm{g}}$ or $\dot{R}$. 
\end{remark}
\begin{lem}
\label{lem:UAV_speed_varying}
If the curvature of the orbit $O$ is bounded by \eqref{eq:orbit_curvature_constraint} then the rate-of-change for the radius $R$ is bounded by 
\begin{equation}
    \label{eq:morph_circle_limit}
    -v + v_{g} \leq \frac{1}{t_j - t_i}(R_j - R_i) \leq v - v_{g}\;.
\end{equation}

\begin{proof}
If orbit $O$ satisfies \eqref{eq:orbit_curvature_constraint} then $\kappa$ is defined by the LHS of \eqref{eq:orbit_curvature_constraint}. 
For \eqref{eq:orbit_curvature_constraint} to be real-valued it is required that $
    v^2 - (\dot{\bm{g}}\cdot\bm{e}_r + \dot{R})^2 \geq 0$.
Moreover, this implies that $ -v \leq \dot{R}  + \dot{\bm{g}}\cdot\bm{e}_r  \leq v$ and 
\begin{align}
 -v -\dot{\bm{g}} \cdot \bm{e}_r &\leq \dot{R} \leq v - \dot{\bm{g}} \cdot \bm{e}_r\label{eq:dotR_inequality_variable}\;.
\end{align}
Since $||\dot {\bm g}|| = v_g$ and $||{\bm e}_r|| = 1$ then  $| \dot{\bm{g}}\cdot\bm{e}_r| \leq v_{g}$. For the lower bound of \eqref{eq:dotR_inequality_variable} to hold for all $\theta$ then $-v + v_g \leq \dot{R}$. Conversely, for the upper bound to hold for all $\theta$ then $\dot{R} \leq v - v_g$. The new inequality can be expressed as 
\begin{equation}
    -v + v_g \leq \dot{R} \leq v - v_g\;.
    \label{eq:dotR_abs_ineq}
\end{equation}
Substituting \eqref{eq:dotR_termsRiRj} into \eqref{eq:dotR_abs_ineq} the bounds in \eqref{eq:morph_circle_limit} are found. 
\end{proof}

\end{lem}
\begin{remark}
    Since Lemma~\ref{lem:UAV_speed_varying} provides a bound on the rate of change of radius of the orbit $\dot{R}$, it creates a limit on the radius of consecutive visibility orbits given a POI trajectory $\bm{g}$. 
\end{remark}
\begin{lem}
\label{lem:lim_curvature}
    Suppose that the system \eqref{eq:dynamics} remains on the orbit $O$ then a  bound on the curvature \eqref{eq:orbit_curvature_constraint} is 
    \begin{equation}
        \kappa_{\rm max} \leq \frac{1}{R}\left(\frac{v_g}{v} + 1\right)^2\;.
        \label{eq:lim_curvature}
    \end{equation}
    \begin{proof}
    The maximum of \eqref{eq:orbit_curvature_constraint} can be found by maximizing the function 
    \begin{equation}
        J = \frac{\left(b\sin(a) \pm \sqrt{1 - (b\cos(a) + c)^2}\right)^2}{(d +vct)\sqrt{1 - (b\cos(a) + c)^2}}\;,
        \label{eq:symbolic_cost}
    \end{equation}
    which is the value of $\kappa$ obtained by substituting \eqref{eq:ang_vel} into \eqref{eq:orbit_curvature_constraint}, using the identities $\bm{g}\cdot\bm{e}_r = v_g\cos a$ and $\bm{g}\cdot\bm{e}_\theta=v_g\sin a$, defining $a=\theta - \gamma$ as the difference between the polar angle and the POI's direction, introducing the speed ratio $b = \frac{v_g}{v}$, defining the ratio between the orbits radial rate and the UAV's speed $c = \frac{\dot{R}}{v}$, the initial radius $d=R_0$, and $R = d + vct$ as the orbit radius at any time $t$.
    The bounds of the maximization are the bounds on the rate of change of the orbit's radius from \eqref{eq:dotR_abs_ineq},
    $ 1  \geq b + c$ and $ -1 \leq - b + c$.
    The Jacobian of \eqref{eq:symbolic_cost} was computed symbolically but is not shown due to space constraints. 
    Figure~\ref{fig:jacobian} illustrates the cost function \eqref{eq:symbolic_cost} for CCW motion with parameters $d = 1$, and $t = 0$ in the dark red to white color scale. The curves where elements of the Jacobian are zero also overlayed in Fig.~\ref{fig:jacobian} as the blue dashed and green lines.
    \begin{figure}[h]
        \centering
        \includegraphics[
        width=0.8\textwidth
        ]{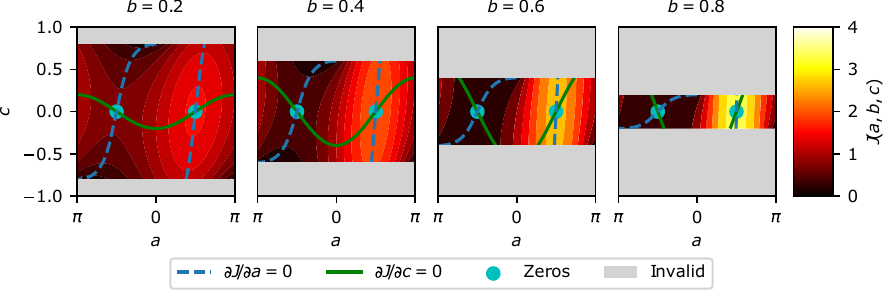}
        \caption{A plot of the cost function \eqref{eq:symbolic_cost} and its Jacobian where $d=1$, $v=1$ and $t=0$ and for varying values of $b$.}
        \label{fig:jacobian}
    \end{figure}
    In Fig.~\ref{fig:jacobian}, two local extrema are observed to always occur at $a = \pm \frac{\pi}{2}$ and $c=0$ shown as cyan dots.
    In Fig.~\ref{fig:jacobian}, the gray regions are invalid according to Lemma~\ref{lem:UAV_speed_varying}.
    For any set of $d, t$, there is an equivalent value of $d', t'=0$ such that $cvt + d = cvt' + d'$; Therefore, evaluating the Jacobian at $t'=0$ and $d'$ is valid for the corresponding $t>0$ and $d$.
    Evaluating the Jacobian at $a = \pm \frac{\pi}{2}$ and $c = 0$ symbolically results in the zero vector and shows that the critical point is independent of $b$, $d$, and $v$. Thus, the function \eqref{eq:symbolic_cost} has local extrema at $[\pm\frac{\pi}{2}, 0]^{\rm T}$ with a value of,
    \begin{equation}
        J\left(\pm \frac{\pi}{2}, 0\right) =\frac{\left(b \pm 1\right)^2}{d}= \frac{\left(v_g \pm v\right)^2}{R_0v^2}\;.
        \label{eq:local_extrema}
    \end{equation}
    Based on an analysis of plots,
    the maximum for CCW motion is at $\frac{\pi}{2}$ and $-\frac{\pi}{2}$ for CW motion. 
    \end{proof}
\end{lem}
\begin{cor}
    Since $v_g\leq v$, then from Lemma~\ref{lem:lim_curvature}, the maximum of the right-hand side of \eqref{eq:lim_curvature} is $\kappa=\frac{4}{R}$. This indicates that a UAV with speed $v$ can track a moving POI with any speed $v_g \leq v$ as long as the orbit radius is four times greater than the minimum radius $R\geq 4r_{\rm min}$. This is useful because it creates a UAV turn radius selection criterion for a mission.
\end{cor}
\begin{remark}
    Lemma~\ref{lem:lim_curvature} demonstrates the surprising fact that the largest curvature occurs when $\dot R = 0$. In other words, a changing radius orbit that is also translating requires less maneuvering than a constant radius orbit translating in the same manner. The point of maximum curvature occurs at $a = \pm \pi/2$ which correspond to the  antipodal points on the orbit that form a line perpendicular to the POI's motion.
\end{remark}

\subsection{Representing Moving View Volumes via an Adaptive Discretization Approach}
\label{sec:adaptive}
VVs for a moving POI remain relatively constant in shape over a road network when the surrounding building geometry also remains relatively unchanged (e.g., when traversing between two spatially uniform structures). However, when one structure ends or changes shape, the VV geometry varies. An adaptive discretization approach chooses an appropriate spacing over the road network for which to compute VVs, such that more VVs are located in regions of greater environmental variability. The adaptive discretization approach relies on quantifying the difference between neighboring VVs $\hat{\alpha}(\bm{g}_i)$ and $\hat{\alpha}(\bm{g}_j)$, computed at two points $\bm{g}_i$ and $\bm{g}_j$, using the difference metric:
\begin{equation}
    d(\hat{\alpha}(\bm{g}_i), \hat{\alpha}(\bm{g}_j)) = \texttt{volume}\left(\hat{\alpha}(\bm{g}_i) \oplus \hat{\alpha}(\bm{g}_j)\right) \;, \label{eq:metric}
\end{equation} 
where $\oplus$ is the exclusive or operator, and \texttt{volume} is a function that returns the volume of the resulting region. Intuitively, this operator returns the volume of the region that is in either $\hat{\alpha}(\bm{g}_i)$ or in $\hat{\alpha}(\bm{g}_j)$ but not in both.

The proposed adaptive approach is inspired by a bisection method.
The approach increases the number of VVs along a road network edge where the VVs are quickly changing by placing a new VV halfway between two VVs.
The approach requires the start and end points of an edge on the road network, $\bm{g}_i$ and $\bm{g}_j$, a set of initial points along the road network, and a constant threshold parameter $d_{\rm cutoff}$ that is repeatedly compared to \eqref{eq:metric}.
The output of the approach is a set of points $\mathcal{P}$ at which the VVs were computed.
The approach recursively calls itself twice if the metric \eqref{eq:metric} is greater than $d_{\rm cutoff}$. The first call updates the endpoint parameter to the middle point, and the second call updates the start point parameter to the middle point (creating two new road segments half the size of the original). 
\begin{figure}[h!]
    \centering
    \includegraphics[width = 0.7\textwidth]{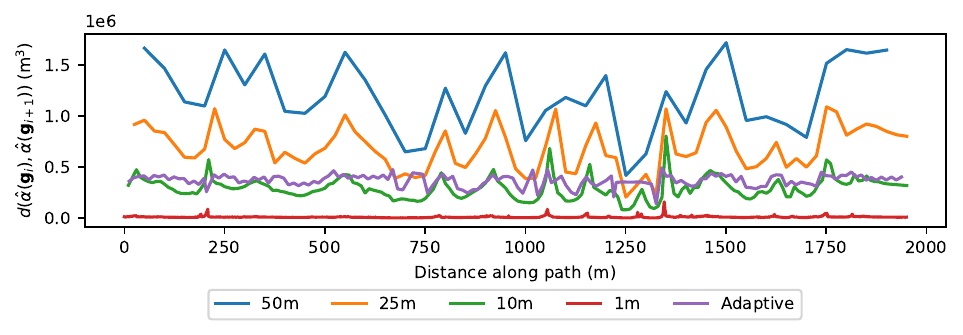}
    \caption{ A plot of the VV similarity metric along a path through an urban environment for fixed distance discretization methods ($50~{\rm m}$ blue, $25~{\rm m}$ orange, $10~{\rm m}$ green, and $1~{\rm m}$ red) and the adaptive discretization approach (purple) with a cutoff at $d_{\rm cutoff} = 500,000$ m$^3$.}
    \label{fig:metric}
\end{figure}

Figure~\ref{fig:metric} plots the distance \eqref{eq:metric} computed along a path through a city to compare the adaptive discretization with several fixed-spacing discretizations with uniform spacings of $1~{\rm m}$, $10~{\rm m}$, $25~\rm{m}$, $50~{\rm m}$. The metric~\eqref{eq:metric} varies greatly for the larger spacings, but the adaptive discretization method maintains the metric at an approximately constant level or below $d_{\rm cutoff} = 500,000$ m$^3$ (about 25\% of the volume of a hemispherical VV with a radius of 100~m). The value for the adaptive discretization is similar to the $10~{\rm m}$ spacing but reduces the spike around $1300~{\rm m}$ along the path. A uniform discretization with $1~{\rm m}$ spacing creates a refined representation of the VVs, but it requires more computation.

\subsection{Locus of Circular Orbits}
\label{sec:circular_orbits}
Once the path of the POI on the road network has been discretized into a series of $m$ points $\mathcal{P} = \{ {\bm g}_0, \ldots, {\bm g}_{m-1} \}$ with corresponding VVs $\hat \alpha({\bm g}_i)$ and times $t_{0:m-1} = \{t_0, \ldots, t_{m-1}\}$, then the locus of orbits is created with radii $R_{0:m-1} = \{R_0, \ldots, R_{m - 1}\}$ with centers corresponding to points in $\mathcal{P}$. These radii define the piecewise-linear time-varying radius $R$ as described in Sec.~\ref{sec:feasibility}. The objective \eqref{eq:optimization_prob} is to maximize the standoff distance to the target. Therefore, the radii $R_i$ are maximized within the VPs (at altitude $h_{\rm UAV}$) 
 $\hat \alpha({\bm g}_i)$ while remaining feasible for the Dubins vehicle.

\begin{figure}[h!]
    \centering
    \includegraphics[width=0.7\textwidth]{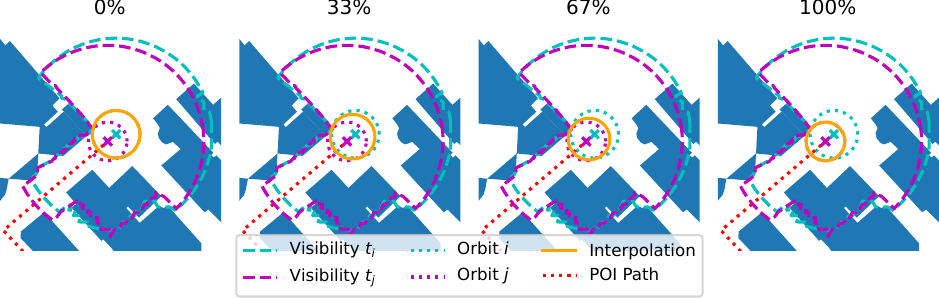}
    \caption{A circular orbit inside the (cyan) visibility polygon morphing (orange) to another visibility polygon (magenta) $10~{\rm m}$ down the future path of the POI (red dotted line).}
    \label{fig:circle_morph}
\end{figure}

The proposed method for determining the radii of the circular orbit is outlined in Algorithm~\ref{alg:circular}. 
\begin{algorithm}[h!]
  \caption{Circular Orbits}
  \begin{algorithmic}[1] 
    \STATEx \hspace{-2.1em}  {\bf function:} \texttt{CircularOrbits}$(\bm{g}(t), \mathcal{P}, t_{0:m}, \kappa_{\rm max}, v, h_{\rm UAV})$
    \STATEx \hspace{-2.1em}   {\bf input:} $\bm{g}(t)$ the POI trajectory, a set $\mathcal{P}$ of $m$ points along the trajectory $\bm{g}(t)$, $t_{0:m - 1}$ the set of times corresponding to the points, $\kappa_{\rm max}$ the maximum curvature of the UAV, $v$ the UAV's speed
    \STATEx \hspace{-2.1em}  {\bf output:} an ordered set of radii $R_{0:m - 1}$ where all points on the orbit can view the corresponding point in $\mathcal{P}$
    \STATE $R_{0:m} \gets \{\texttt{MaxRadius}(\alpha(\mathcal{P}_i), h_{\rm UAV})~{\rm for~all}~\mathcal{P}_i \in \mathcal{P}\}$\label{alg:circular:bisection}
    \FOR {$i \in \rangeset{m - 1}$}  \label{alg:circular:radius+0}
        \STATE $\dot{R} \gets ({R_{i + 1} - R_{i}})/({t_{i + 1} - t_{i}})$
        \IF {$\dot{R} > v - v_{g}$}
            \STATE $R_{i + 1} \gets (v - v_{g})(t_{i + 1} - t_i)$
        \ENDIF
    \ENDFOR\label{alg:circular:radius+1}
    \FOR {$i \in \{m - 1, \ldots, 0\}$} \label{alg:circular:radius-0}
        \STATE $\dot{R} \gets ( {R_{i + 1} - R_{i}})/({t_{i + 1} - t_{i}})$ 
        \IF {$\dot{R} < v_{g} - v$}
            \STATE $R_{i} \gets (v_{g} - v)(t_{i + 1} - t_{i})$
        \ENDIF
    \ENDFOR \label{alg:circular:radius-1}
    
    \FOR {$i \in \rangeset{m - 1}$}
        \IF {$R_i > \frac{1}{\kappa_{\rm max}}\left(\frac{v_g}{v} + 1\right)^2$} \label{alg:circular:curvature}
            \STATE {\bf return} \texttt{Infeasible}
        \ENDIF
    \ENDFOR
  \end{algorithmic}
\label{alg:circular}
\end{algorithm}
The algorithm first uses the bisection method \texttt{MaxRadius} to find the largest circle centered on $\bm{g}_i$ that can fit inside each VV $\hat \alpha({\bm g}_i)$ at height $h_{\rm UAV}$ (Alg.~\ref{alg:circular}, Line \ref{alg:circular:bisection}). The process of finding the largest VO is similar to the maximum inscribed circle problem 
\cite{Huang2021}.
The bisection method starts with the minimum radius zero and the maximum sensing distance $d_{\rm max}$ and iteratively increases the minimum and decreases the maximum radius until they are arbitrarily close. 
The results is a circle that touches the VP at least once.
Then, to ensure that the UAV can fly along the morphing orbit, the rate of change of the orbits is checked with to Lemma~\ref{lem:UAV_speed_varying}. 
The first for loop (Alg.~\ref{alg:circular}, Lines~\ref{alg:circular:radius+0}--\ref{alg:circular:radius+1}) iterates forward through the orbits, enforcing the upper bound from \eqref{eq:morph_circle_limit}. 
The second for loop (Alg.~\ref{alg:circular}, Lines~\ref{alg:circular:radius-0}--\ref{alg:circular:radius-1}) iterates backwards, enforcing the lower bound from \eqref{eq:morph_circle_limit}. 
Then, the resulting morphing orbit is checked for curvature constant violations (Alg.~\ref{alg:circular}, Line~\ref{alg:circular:curvature}) with Lemma~\ref{lem:lim_curvature}. If the check for the curvature of the orbit fails, the path $\bm{g}$ is deemed infeasible for system \eqref{eq:dynamics}--\eqref{eq:dynamics_constraint} with speed $v$ and curvature bound $\kappa_{\rm max}$. 
An example of the resulting circular orbits can be seen in Fig.~\ref{fig:circle_morph}.
Where the orbit is morphing from the $i$th magenta orbit inside the $i$th VV to the $j$th cyan orbit inside the $j$th VV. The $i$th and $j$th VVs are calculated along the POI's path, Xs of the corresponding color. The blue footprints of the buildings are shown in Fig~\ref{fig:circle_morph} and the VPs are calculated at $h_{\rm UAV} = 50$~m. Figure~\ref{fig:circle_morph} illustrates that circular orbits restrict the UAV to a small percentage of the visibility polygon for some environments. Future work may consider more general visibility orbit geometries.

\section{Feedback Controller}
\label{sec:feed_controller}

This section describes a steering controller that drives the Dubins vehicle, system \eqref{eq:dynamics}, onto the orbits created in Sec.~\ref{sec:circular_orbits}. First, Sec.~\ref{sec:desired_heading} describes the construction of a constant magnitude vector field using a framework from \cite{Goncalves2010}. The vector field defines a desired heading guidance command for the vehicle. Next, Sec.~\ref{sec:heading_rate} derives an expression for the heading-rate along the vector field. Lastly, Sec.~\ref{sec:steering_controller} proposes a steering controller $u_\psi$ with three terms: (i) a proportional term that forces the UAV's heading to the desired guidance heading, (ii) a feed-forward term that uses the derived heading-rate, and (iii) a term to ensure that the error between the orbit and the UAV approaches zero. 
\subsection{Vector Field Guidance for the Desired Heading}
\label{sec:desired_heading}
\begin{figure}[t]
    \centering
    \includegraphics[width=0.8\textwidth]{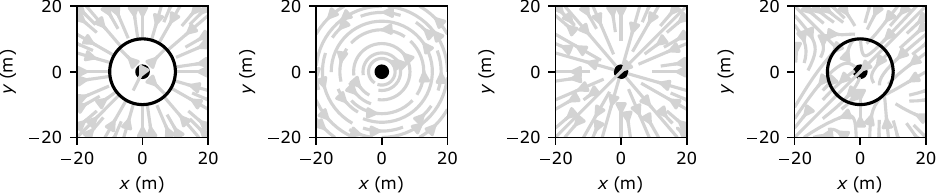}
    \caption{The parts of a time-varying circulation vector field in \eqref{eq:general_field} for counter-clockwise circulation: attraction, feedforward, circulation, and a combination of the attraction and feedforward, from left to right.
    }
    \label{fig:three_fields}
\end{figure}

\begin{figure}[t]
    \centering
    \includegraphics[width=0.4\textwidth]{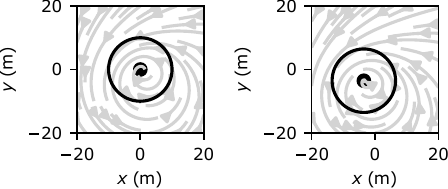}
    \caption{
    The vector field \eqref{eq:constant_vector_field} (gray), for an orbit (black) moving to the bottom left after $1$~s.
    }
    \label{fig:field_timevarying}
\end{figure}

The authors in \cite{Goncalves2010} created a generic framework for finding a vector field to attract a single integrator model of a system with state $\mathbb{R}^n$ to a time-varying closed trajectory. This work uses that framework to find a specific realization of that vector field for a constant speed single integrator model in $\mathbb{R}^2$ to follow the orbit $O$. The resulting vector field generates guidance commands for a steering controller $u_\psi$. 
Consider the single integrator model
\begin{equation}
    \dot{\bm{\xi}} = \bm{u}\;,
    \label{eq:single_integrator}
\end{equation}
where $\bm{\xi} = [\xi_x~\xi_y]^\T \in \mathbb{R}^2$ is the position of a particle with velocity control ${\bm u} \in [u_x,~u_y]^\T$. The method in \cite{Goncalves2010} relies on encoding the desired trajectory into a series of $n - 1$ functions denoted $a_i(\bm{\xi}, t) \in \mathbb{R}$ that are zero only along the desired trajectory and have bounded second derivatives. The vector field in \cite{Goncalves2010} has the form
\begin{equation}
    \label{eq:general_field}
    \bm{u} = -\Phi \nabla V_1 + H\prescript{\star}{}{\left(\wedge_{i = 1}^{n-1} \nabla a_i\right)} - \bm{M}^{-1}\dot{\bm{\sigma}}\;,
\end{equation}
where $\Phi(\bm{\xi}, t) \in \mathbb{R}$ is a strictly positive function for ${\bm \xi} \notin O$, $V_1(a_1, \ldots,a_{n-1}) \in \mathbb{R}^+$ is a strictly positive function that depends on the $a_i$ described earlier, $H(\bm{\xi}, t) \in \mathbb{R}^+\backslash\{0\}$ or $H(\bm{\xi}, t) \in \mathbb{R}^-\backslash\{0\}$ is either a strictly positive or a strictly  negative function defining CCW or CW traversal of the trajectory, $\wedge_{i = 0}^{n-1} \nabla a_i$ is the wedge product \cite{darling1994differential} between the gradient of the functions $a_i$ creating $n${\rm-1}-blades, $\prescript{\star}{}{(\cdot)}$ is the Hodge star \cite{darling1994differential} which maps $n${\rm-1}-blades to {\rm1}-blades or vectors, $\bm{M}$ is an invertible $n\times n$ matrix where the first $n-1$ rows are $\nabla a_i^\T$ and the last row is $\prescript{\star}{}{\left(\wedge_{i = 1}^{n-1} \nabla a_i\right)}^{\rm T}$, and the vector $\dot{\bm{\sigma}}$ is $\partial a_i/\partial t$ for the first $n-1$ rows and $0$ for the last row.
In the controller \eqref{eq:general_field}, the $\Phi\nabla V$ term is the direction to the orbit $O$. 
The controller \eqref{eq:general_field} differs from the notation used in \cite{Goncalves2010}; We have elected to use the Hodge star operator to conform to our convention of using column vectors, as opposed to row vectors, in the wedge product.
The next term, $H\prescript{\star}{}{\left(\wedge_{i = 1}^{n-1} \nabla a_i\right)}$, creates a vector orthogonal to $\Phi\nabla V$ for circulation on the orbit $O$. The final term $\bm{M}^{-1}\dot{\bm{\sigma}}$ is a feed-forward term ensuring the UAV remains on the time-varying orbit $O$. To implement a controller from the framework in \cite{Goncalves2010}, consider the function
\begin{equation}
    a_1 = \arctan\left(\beta\left(\sqrt{(\xi_x - g_x)^2 + (\xi_y - g_y)^2} - R\right)\right)\;,
    \label{eq:bowl}
\end{equation}
where $\beta$ is a non-zero positive scalar controlling attraction to the orbit, and $a_1$ is zero on the orbit $O$ for all ${\bm \xi} \in O$, and $g_x, g_y$ and $R$ are time-varying terms as introduced earlier. 
The partial time derivative of the function \eqref{eq:bowl} is
\begin{equation}
    \frac{\partial a_1}{\partial t} = \frac{\beta}{1 + \beta^2\left(||\bm{\xi} - \bm{g}|| - R\right)^2}\left(-\dot{{R}} - \frac{\dot{g}_x(\xi_x - g_x) + \dot{g}_y(\xi_y - g_y)}{||\bm{\xi} - \bm{g}||}\right)\;.
    \label{eq:bowl_dt}
\end{equation}
The potential function
$
    V_1 = \frac{1}{2} a_1^2\notag
$
creates a potential bowl with the bottom of the bowl along the orbit. 
The feed-forward term $\bm{M}^{-1}\dot{\bm{\sigma}}$ is created by finding the $\bm{M}$ matrix, which involves the gradient  
\begin{equation}
    \nabla a_1 = 
\begin{bmatrix}
    \frac{\partial a_1}{\partial \xi_x} \\
    \frac{\partial a_1}{\partial \xi_y} 
\end{bmatrix}
    =
    \frac{\beta}{r\left(1 + \beta^2(r- R)^2\right)}\begin{bmatrix}
        \xi_x - g_x\\
        \xi_y - g_y
    \end{bmatrix}\notag
\end{equation}
on the first row, where  $r = ||\bm{\xi} - \bm{g}||$ is the distance between the POI $\bm{g}$ and the inertial position $\bm{\xi}$, and the
circulation term
\begin{equation}
    \prescript{\star}{}{\left(\wedge_{i=1}^{1} \nabla a_i\right)} = \prescript{\star}{}{\left(\nabla a_1\right)} = \frac{\beta}{r \left(1 + \beta^2(r - R)^2\right)}\begin{bmatrix}
        g_y - \xi_y\\
        \xi_x - g_x
    \end{bmatrix}\;,
    \label{eq:wedgecircular}
\end{equation}
on the second (i.e., last) row where the wedge product for one vector is an identity operation $\wedge^1_{i=1}\nabla a_i=\nabla a_1$, and the Hodge star in $\mathbb{R}^2$ transforms the directions $\prescript{\star}{}{\bm{e}_x} = \bm{e}_y$ and $\prescript{\star}{}{\bm{e}_y}=-\bm{e}_x$.
The resulting matrix $\bm{M}$ is 
\begin{equation}
    \bm{M} = \frac{\beta}{r\left(1 + \beta^2(r - R)^2\right)}\begin{bmatrix}
        \xi_x - g_x & \xi_y - g_y\\
        g_y - \xi_y & \xi_x - g_x
    \end{bmatrix}\;.\notag
\end{equation}
For all $r\neq 0$ the matrix $\bm{M}$ is an orthonormal basis multiplied by a function. Since the determinant is non-zero,
\begin{align}
    \det(\bm{M}) &= \frac{\beta^2}{\left(1 + \beta^2(r - R)^2\right)^2}\left(\frac{(\xi_x - g_x)^2 + (\xi_y - g_y)^2}{r^2}\right) \notag\\& = \frac{\beta^2}{\left(1 + \beta^2(r - R)^2\right)^2}\;,\notag
\end{align}
the feed-forward term is
\begin{equation}
    \bm{M}^{-1}\dot{\bm{\sigma}} = \frac{1 +\beta^2 (r - R)^2}{\beta r}\begin{bmatrix}
        \xi_x - g_x & g_y - \xi_y\\
        \xi_y - g_y & \xi_x - g_x
    \end{bmatrix}
    \begin{bmatrix}
        \frac{\partial a_1}{\partial t} \\
        0
    \end{bmatrix}\;,\notag
\end{equation}
where the vector $\dot{\bm{\sigma}}=[\partial a_1/\partial t, 0]^\T$ \cite{Goncalves2010}. The expression for the feed-forward term can be simplified using the polar coordinate system about the point $\bm{g}$,
\begin{align}
    \bm{M}^{-1}\dot{\bm{\sigma}} &= 
    \frac{\frac{\partial a_1}{\partial t}\left(1 + \beta^2(r - R)^2\right)}{\beta r}
    \begin{bmatrix}
        \xi_x - g_x\\
        \xi_y - g_y
    \end{bmatrix}\notag
    \\
    &=\frac{\partial a_1}{\partial t}\frac{1}{\beta}\left(1 + \beta^2(r - R)^2\right) \bm{e}_r\;,
    \label{eq:ff_partial}
\end{align}
where $\bm{e}_r$ and $\bm{e}_\theta$ are from Definition~\ref{def:polar}. By substituting \eqref{eq:bowl_dt} into \eqref{eq:ff_partial}, and noting that ${\bm e}_r = 1/r[(\xi_x-g_x)~(\xi_y-g_y)]^{\rm T}$, the feed forwarded term becomes
\begin{align}
   \bm{M}^{-1}\dot{\bm{\sigma}} 
&= \left[ \frac{\beta}{1 + \beta^2\left(||\bm{\xi} - \bm{g}|| - R\right)^2}\left(-\dot{{R}} - \frac{\dot{g}_x(\xi_x - g_x) + \dot{g}_y(\xi_y - g_y)}{||\bm{\xi} - \bm{g}||}\right) \right] \left[ \frac{1}{\beta}\left(1 + \beta^2(r - R)^2\right) \bm{e}_r 
\right] \notag\\ 
&=
\left(-\dot{{R}} - \frac{\dot{g}_x(\xi_x - g_x) + \dot{g}_y(\xi_y - g_y)}{||\bm{\xi} - \bm{g}||}\right) \bm{e}_r \notag \\ 
   &= \left(-\dot{R}-\dot{\bm{g}}\cdot\bm{e}_r\right)\bm{e}_r
    \label{eq:feedforward}\;.
\end{align}
The gradient term in \eqref{eq:general_field} is 
\begin{align}
    \nabla V_1 &= a_1 \frac{\beta}{r\left(1 + \beta^2(r  - R)^2\right)} \begin{bmatrix}
        \xi_x - g_x\\
        \xi_y - g_y
    \end{bmatrix}\notag\\
    &= \frac{\beta \arctan\left(\beta(r - R)\right)}{1 + \beta^2(r-R)^2} \bm{e}_r\;
    \label{eq:gradient}
\end{align} 
and attracts the single integrator to the orbit.
As will be shown momentarily, the term $H\prescript{\star}{}{\left(\wedge_{i = 1}^{n-1} \nabla a_i\right)}$ in the controller ${\bm u}$ given by \eqref{eq:general_field} is entirely along the ${\bm e}_\theta$ direction. Thus,
\begin{equation}
    {\bm u} \cdot {\bm e}_r = (-\Phi \nabla V + H\prescript{\star}{}{\left(\wedge_{i = 1}^{n-1} \nabla a_i\right)} - \bm{M}^{-1}\dot{\bm{\sigma}}) \cdot {\bm e}_r =  (-\Phi \nabla V_1  - \bm{M}^{-1}\dot{\bm{\sigma}}) \cdot {\bm e}_r \;.
\end{equation}
To design a controller with constant magnitude $||\bm{u}||=v$, a necessary condition is that the radial velocity component is bounded to less than the speed of the system \eqref{eq:dynamics},
\begin{equation}
    \left|\left|\left(\Phi\nabla V - \bm{M}^{-1}\dot{\bm \sigma}\right)\cdot \bm{e}_r\right|\right| \leq v\;.
    \label{eq:radial_vel_bound}
\end{equation}
A constant magnitude controller is desired, and not considered in \cite{Goncalves2010} because trajectories $\bm{\xi}(t)$ can be used as guidance for system \eqref{eq:dynamics}; Whereas, \cite{Goncalves2010} created the controller for a single integrator.
The maximum magnitude of \eqref{eq:gradient} is
\begin{equation}
    \frac{\pi}{2}\frac{\beta}{1 + \beta^2(r-R)^2}\;,
    \label{eq:new_bounded_func}
\end{equation} 
and the maximum magnitude of \eqref{eq:feedforward} is $v_g + |\dot{R}|$. By combining the maximum magnitude of \eqref{eq:feedforward} and the inverse of \eqref{eq:new_bounded_func} the attraction function $\Phi$ appearing in \eqref{eq:general_field} is selected as 
\begin{equation}
    \Phi = \left(v - v_g -\left|\dot{R}\right|\right)\frac{2}{\pi}\frac{\left(1 + \beta^2(r  - R)^2\right)}{\beta}\;,
    \label{eq:phi_expr}
\end{equation}
which is always positive when Lemma~\ref{lem:UAV_speed_varying} is satisfied.
The bound \eqref{eq:radial_vel_bound} considering \eqref{eq:feedforward}, \eqref{eq:phi_expr}, \eqref{eq:gradient} is
\begin{equation}
    \left|\left(v - v_g -\left|\dot{R}\right|\right)\frac{2}{\pi}\arctan\left(\beta (r - R)\right) - \left(-\dot{\bm{g}}\cdot\bm{e}_r - \dot{R}\right)\right|||\bm{e}_r|| \leq v\;,\notag
\end{equation}
and is satisfied for all $\bm{\xi} \in \mathbb{R}^2$. When \eqref{eq:phi_expr} is multiplied by the gradient \eqref{eq:gradient},
\begin{equation}
    \Phi \nabla V_1 = \left(v - v_g -\left|\dot{R}\right|\right)\frac{2}{\pi}\arctan\left(\beta (r - R)\right)\bm{e}_r\;,
    \label{eq:grad_phi}
\end{equation}
the attractive velocity is found. The scalar component of \eqref{eq:grad_phi} is denoted  
\begin{equation}
    ||\Phi \nabla V_1|| = \Phi' = \left(v - v_g -\left|\dot{R}\right|\right)\frac{2}{\pi}\arctan\left(\beta (r - R)\right)\; 
    \label{eq:phi_simp}
\end{equation}
for the simplification of future notation.
The vector $\prescript{\star}{}{\left(\nabla\wedge_{i=1}^1a_i\right)}$ in \eqref{eq:wedgecircular} can be written as 
\begin{equation}
    \prescript{\star}{}{\left(\wedge_{i=1}^1a_i\right)} =  \frac{\beta}{\left(1 + \beta^2(r  - R)^2\right)}\bm{e}_\theta\;.
    \label{eq:wedge_r2}
\end{equation}
Solve for $H$ so that the magnitude of the controller \eqref{eq:general_field}---made of \eqref{eq:feedforward}, \eqref{eq:grad_phi}, \eqref{eq:phi_simp}, and \eqref{eq:wedge_r2}---is  the speed of system \eqref{eq:dynamics}
\begin{align}
    v^2 = ||\bm{u}||^2 &= \left|\left|-\Phi'\bm{e}_r + \left(\dot{R} + \dot{\bm{g}}\cdot \bm{e}_r\right)\bm{e}_r + H\frac{\beta}{\left(1 + \beta^2(r  - R)^2\right)}\bm{e}_\theta\right|\right|^2\notag\\
    H &= \pm\frac{1 + \beta^2(r  - R)^2}{\beta}\sqrt{v^2 - \left(-\Phi' + \dot{R} + \dot{\bm{g}}\cdot \bm{e}_r\right)^2}\;,
    \label{eq:circulate}
\end{align}
where the ``+'' and ``-'' solutions are CCW and CW rotation, respectively.
Expanding \eqref{eq:general_field} with \eqref{eq:feedforward}, \eqref{eq:phi_simp}, \eqref{eq:wedge_r2}, and \eqref{eq:circulate},
\begin{align}
    \bm{u} & = \left(-\Phi' + \dot{R} + \dot{\bm{g}}\cdot \bm{e}_r\right) \bm{e}_r 
    \pm  \left(\sqrt{v^2 - \left(-\Phi' + \dot{R} + \dot{\bm{g}}\cdot \bm{e}_r\right)^2}\right) \bm{e}_\theta\label{eq:constant_vector_field}\;.
\end{align}
The vector field \eqref{eq:constant_vector_field} consists of three parts shown in Fig.~\ref{fig:three_fields}: (i) $-\Phi'\bm{e}_r$ radial attraction, (ii) $(\dot{R} + \dot{\bm{g}} \cdot \bm{e}_r)\bm{e}_r$ feed-forward, 
and (iii) circulation $\left(\sqrt{v^2 - \left(-\Phi' + \dot{R} + \dot{\bm{g}}\cdot \bm{e}_r\right)^2}\right) \bm{e}_\theta$. 
The combination of the radial attraction and the feed-forward terms $(-\Phi' + \dot{R} + \dot{\bm{g}}\cdot\bm{e}_r)\bm{e}_r$ are shown in Fig.~\ref{fig:three_fields} to compare the magnitude of the two radial fields.
Trajectories of a particle moving in $\mathbb{R}^2$ using the controller $\bm{u}$ are visualized in Fig.~\ref{fig:field_timevarying}. The vector field was created with a POI moving at $v_g=5$~m/s with heading $\gamma=-3\pi/4$~rad with $\beta=0.1$ and the UAV moving at $10$~m/s. The velocity streamlines do not arrive tangentially at the orbit because the orbit is translating to the bottom right corner---the UAV will not remain on the orbit in the next time step if the streamlines are tangent to the orbit.
The controller \eqref{eq:constant_vector_field} can be expanded using \eqref{eq:phi_simp} to
\begin{align}
    \bm{u} &= \left(\left(-v + v_g + \left|\dot{R}\right|\right)\frac{2}{\pi}\arctan(\beta (r - R)) + \dot{\bm{g}} \cdot \bm{e}_r + \dot{R}\right)\bm{e}_r\notag\\&\pm \left(\sqrt{v^2 - \left(\left(-v + v_g + \left|\dot{R}\right|\right)\frac{2}{\pi}\arctan\left(\beta(r - R)\right) + \dot{\bm{g}} \cdot \bm{e}_r + \dot{R}\right)^2}\right) \bm{e}_\theta\;.
    \label{eq:controller_expanded}
\end{align}
This controller builds upon the framework in \cite{Goncalves2010} by finding a specific implementation of the generalized controller for a constant velocity single integrator model. To implement the controller, this work found functions $ a_1$, $\Phi$, and $H$ that created a constant magnitude vector field that is tailored towards tracking the reference orbits introduced in Sec.~\ref{sec:polygons}.
\subsection{Heading Rate}
\label{sec:heading_rate}
\begin{figure}[h!]
    \centering
    \includegraphics[width=0.7\textwidth]{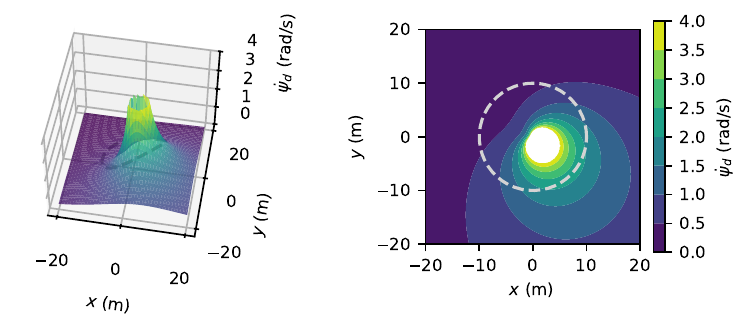}
    \caption{A surface plot and contour plots of $\dot{\psi}_d$ where the middle is cropped at $\dot{\psi}_d = 4$~rad/s.}
    \label{fig:curvature}
\end{figure}

The vector field  \eqref{eq:controller_expanded}  written in terms of polar coordinates $r$ and $\theta$ is rotated to the rectangular coordinates with,
\begin{align}
        \begin{bmatrix}
        u_x \\
        u_y
    \end{bmatrix}
     =\begin{bmatrix}
        \cos\theta & -\sin\theta\\
        \sin\theta & \cos\theta
    \end{bmatrix}
    \begin{bmatrix}
        u_r \\
        u_\theta
    \end{bmatrix}\;.
    \label{eq:coordinate_rotation}
\end{align}
Since the magnitude of the vector field is the same as the Dubins vehicle speed, \eqref{eq:orbit_curvature_constraint} rewritten in $xy$ coordinates is 
\begin{equation}
    \bm{u} = \begin{bmatrix}
        v \cos\psi_d \\
        v \sin\psi_d
    \end{bmatrix}\;, \label{eq:controller_speed_heading}
\end{equation}
where $\psi_d$ is the direction of the vector field 
\begin{equation}
    \psi_d= \arctan \left(\frac{u_y}{u_x} \right)\;,
    \label{eq:psi_d}
\end{equation}
which is a function of the current $xy$ position and time $t$.
If the the UAV starts with initial condition $\bm{q}_0 = [q_x, q_y, \psi_d([q_x, q_y]^{\rm T}, 0)]^\T$
then the system \eqref{eq:dynamics} with an unbounded turn rate can flow along the vector field $\bm{u}$ with the substitution $u_\psi = \dot{\psi}_d = v\kappa_s$
the expression of curvature from \eqref{eq:curvature_2d}. Lemma~\ref{lem:curvature_polar} derives an expression for the curvature of a constant velocity curve in polar coordinates. By using the substitution $\dot{\bm{x}}_\mathcal{I}=\dot{\bm{\xi}}=\bm{u}$ with equation \eqref{eq:curvature_polar} from Lemma~\ref{lem:curvature_polar},
\begin{equation}
\dot\psi=-\frac{\frac{\rm d}{{\rm d}t}\left(\bm{u}\cdot\bm{e}_r\right)}{\bm{u}\cdot{\bm e}_\theta} + \dot{\theta}\;.
    \label{eq:desired_control_rate}
\end{equation}
The Dubins vehicle can use this vector field as a controller with $u_\psi = \dot{\psi}_d$ as the control if (i) $|\dot{\psi}_d| \leq u_{\psi, {\rm max}}$ the rate of change of the direction of the vector field is less than the maximum turn rate, and (ii) the UAV's initial heading is $q_{\psi, 0} = \psi_d(t_0)$.
The term in the numerator of \eqref{eq:desired_control_rate} can be found by differentiating the $r$ component of \eqref{eq:constant_vector_field},
\begin{equation}
    \frac{\rm d}{{\rm d}t}\left(\bm{u}\cdot\bm{e}_r\right)= \frac{\rm d}{{\rm d}t} \left(-\Phi' + \dot{R} + \dot{\bm{g}}\cdot \bm{e}_r\right) = -\dot{\Phi}'  -\dot{\theta}\dot{\bm{g}}\cdot \bm{e}_\theta\;,
    \label{eq:controller_r_deriv}
\end{equation}
where $\ddot{\bm{g}} = 0$ since the POI has a constant velocity, and $\ddot{R} = 0$ since the radius changes linearly. The derivative of \eqref{eq:phi_simp} is 
\begin{equation}
   \dot{\Phi}' =  \frac{\rm d}{{\rm d}t} \left(v - v_g -\left|\dot{R}\right|\right)\frac{2}{\pi}\arctan\left(\beta (r - R)\right) = \left(v - v_g -\left|\dot{R}\right|\right)\frac{2}{\pi}\frac{\beta (\dot{r}-\dot{R})}{1 + \beta^2(r-R)^2}\;.
   \label{eq:phi_dot_exp}
\end{equation} 
In \eqref{eq:phi_dot_exp}, the derivative of the polar radius is 
\begin{align}
    \dot{r} &= \frac{\rm d}{{\rm d}t}\sqrt{(\xi_x - g_x)^2 + (\xi_y - g_y)^2}\notag\\ &= \frac{(\dot{\xi}_x - \dot{g}_x)(\xi_x - g_x) + (\dot{\xi}_y - \dot{g}_y)(\xi_y - g_y)}{\sqrt{(\xi_x - g_x)^2 + (\xi_y - g_y)}}\;,
\end{align}
which can be simplified by substituting in \eqref{eq:single_integrator}, \eqref{eq:poi_velocity}, and the polar direction $\bm{e}_r$ to give 
$
    \dot{r}  = (\bm{u} - \dot{\bm{g}}) \cdot \bm{e}_r
    = u_r - \dot{\bm{g}} \cdot \bm{e}_r
    $.
Substituting \eqref{eq:constant_vector_field} further simplifies this expression to
\begin{equation}
\dot{r}=-\Phi' + \dot{R}\label{eq:controller_rdot}\;.
\end{equation}
The formulation \eqref{eq:phi_dot_exp} can be simplified using \eqref{eq:controller_rdot} and \eqref{eq:phi_simp} to
\begin{align}
    \dot{\Phi}' 
 &= 
    \left(v - v_g -\left|\dot{R}\right|\right)\frac{2}{\pi}\frac{\beta ( -[ \left(v - v_g -\left|\dot{R}\right|\right)\frac{2}{\pi}\arctan\left(\beta (r - R)\right)] )}{1 + \beta^2(r-R)^2}
    \notag \\         
    &= -\frac{4}{\pi^2}\arctan\left(\beta (r - R)\right)\left( \frac{\beta\left(v - v_g -\left|\dot{R}\right|\right)^2}{1 + \beta^2(r-R)^2}\right)\;.
   \label{eq:phi_p_dot}
\end{align}
In \eqref{eq:controller_r_deriv}, the derivative of the polar angle is
\begin{align}
    \dot{\theta} &= \frac{{\rm d}}{{\rm d} t}\arctan\left(\frac{\xi_y - g_y}{\xi_x - g_x}\right)\notag\\
    &=\frac{(\dot{\xi}_y - \dot{g}_y)(\xi_x - g_x) - (\dot{\xi}_x - \dot{g}_x)(\xi_y - g_y)}{r^2}\;,\notag
\end{align}
which is simplified by substituting in \eqref{eq:single_integrator}, \eqref{eq:poi_velocity}, and the polar direction $\bm{e}_\theta$,
\begin{align}
    \dot{\theta}&=\frac{(\bm{u} - \dot{\bm{g}})\cdot \bm{e}_{\theta}}{r}=\frac{u_\theta - \dot{\bm{g}}\cdot \bm{e}_\theta}{r}
    \label{eq:polar_angle_dot}\;.
\end{align}
Finally, by substituting \eqref{eq:polar_angle_dot}, and \eqref{eq:controller_r_deriv} into \eqref{eq:desired_control_rate},
\begin{align}
    \dot{\psi}_d 
    &= \frac{\dot{\Phi}' - \dot{\theta} \dot{\bm{g}}\cdot{ {\bm e}_{\theta}}}{u_\theta} + \dot{\theta}
  = \frac{\dot{\Phi}'}{u_\theta} + \dot{\theta} \frac{r}{u_\theta} \left(\frac{u_\theta - \dot{\bm{g}} \cdot \bm{e}_\theta}{r}\right)\;.\notag
\end{align}
This expression is transformed using the definition of $\dot{\theta}$ in \eqref{eq:polar_angle_dot} to give the desired heading rate,
\begin{equation}
    \dot{\psi}_d = \frac{1}{u_\theta} \left(\dot{\Phi}' + r\dot{\theta}^2\right)\;.
    \label{eq:dot_psi_d}
\end{equation}
A plot of $\dot{\psi}_d$ is shown in Fig.~\ref{fig:curvature}
where the UAV's speed is $10$~m/s, the POI moves with speed $5$~m/s in direction $\frac{-3\pi}{4}$, the gain $\beta=0.1$, and  $t=0$~s. The value of $\dot{\psi}_d$ is small for all values outside of the orbit's radius $10$~m, but there is a singularity where $\dot{\psi}_d$ goes to infinity at $r=0$~m. For large values of $\beta$ the sign is not strictly positive or negative.

\subsection{Attraction to the Orbit}
\label{sec:steering_controller}
If the UAV is initialized with a heading $q_{\psi, 0} \in \mathbb{S}$ that is not aligned with the vector field ${\bm u}$ then a controller is needed to drive it towards $\psi_d$.
The following steering controller is proposed:
\begin{align}
    u_\psi & = k_\psi\left(\psi_d - q_\psi\right) + \dot{\psi}_d |
    \left(\left(\bm{g}\cdot\bm{e}_r + \dot{R}\right)\frac{1-\cos(q_\psi - \psi_d)}{q_\psi - \psi_d}
    + \frac{u_\theta\sin(q_\psi - \psi_d)}{q_\psi - \psi_d}\right)\frac{\beta\frac{2}{\pi}\arctan\left(\beta(r - R)\right)}{1 + \beta^2(r - R)^2}\;,
    \label{eq:controller_proportional}
\end{align}
where $\dot{\psi}_d$ is the rate of change of the heading for a trajectory flowing along the vector field $\bm{u}$ for $\bm{\xi}=[q_x, q_y]^{\rm T}$, $k_\psi$ is a positive proportional gain, and the final term (derived in the next lemma) is required for Lyapunov-like stability \cite{slotine1991nonlinear}.

\begin{cor}
    \label{cor:barbalts}
    From \cite[p. 125]{slotine1991nonlinear}, a corollary of Barbalat's lemma is the following: If a scalar function $V(\bm{q}, t)$ satisfies, (i) $V(\bm{q}, t)$ is lower bounded, (ii) $\dot{V}(\bm{q}, t)$ is negative semi-definite, and (iii) $\dot{V}(\bm{q}, t)$ is uniformly continuous in time,
    then $\dot{V}(\bm{q}, t)\rightarrow0$ as $t\rightarrow\infty$.
\end{cor}
\begin{lem}
\label{lem:controller}
The system \eqref{eq:dynamics} is locally asymptotically attracted to an orbit $O$, satisfying Theorem~\ref{thm:orbit} using controller \eqref{eq:controller_proportional} if the maximum turn rate bound \eqref{eq:dynamics_constraint} is not considered and the gain $k_\psi$ satisfies
\begin{equation}
    k_\psi > v\beta\frac{4}{\pi^2}\epsilon_{\rm max}\;,
    \label{eq:gain_bounds}
\end{equation}
where $\epsilon_{\rm max} \approx0.31483$ (i.e., the maximum of $(\arctan x) / (1 + x^2)$).
\begin{proof}
Consider the scalar function 
\begin{equation}
    V_2 = \frac{1}{\pi}\arctan^2 (\beta\tilde{r}) + \frac{1}{2}\tilde{\psi}^2\;,
    \label{eq:lyapunov}
\end{equation}
where $\tilde{r} = r - R$ is the error between the UAV's radial position and orbit $O$. The heading error $\tilde{\psi}= q_\psi - \psi_d$ is the error between the current and desired headings.
The angular error $\tilde{\psi} \in \mathbb{S}$ is the smaller of the two distances from $\psi_d$ to $q_\psi$, traveling CW or CCW in $\mathbb{S}$. The function \eqref{eq:lyapunov} has a lower bound of zero when $\tilde{\psi}=0$ and $\tilde{r}=0$.
The derivative of \eqref{eq:lyapunov} is
\begin{equation}
    \dot{V}_2 = \frac{\dot{\tilde{r}}\beta\frac{2}{\pi}\arctan{(\beta\tilde{r})}}{1 + \beta^2\tilde{r}^2} + \tilde{\psi}(u_\psi - \dot{\psi}_d)\;,
    \label{eq:lyapunov_dot}
\end{equation}
where $\dot{\tilde{r}}$ is
\begin{align}
    \dot{\tilde{r}} & = \dot{r} - \dot{R}
    = v\cos q_\psi\cos\theta + v\sin q_\psi \sin\theta - \dot{\bm{g}}\cdot\bm{e}_r- \dot{R}\;. \label{eq:r_tilde_a}
\end{align}
The radius rate $\dot{R}$ is constant along the linear path segment of Definition~\ref{def:poi}. Substituting $\tilde{\psi} = q_\psi - \psi_d$ into \eqref{eq:r_tilde_a},
\begin{align}
    \dot{\tilde{r}}
    & = v\cos\theta(\cos\psi_d \cos\tilde{\psi} - \sin\psi_d\sin\tilde{\psi})+ v\sin\theta(\sin\psi_d\cos\tilde{\psi} + \sin\tilde{\psi}\cos\psi_d) - \dot{\bm{g}}\cdot\bm{e}_r- \dot{R}\;. \label{eq:r_tilde_b}
\end{align}
The direction of the vector field and its constant magnitude are expressed in \eqref{eq:controller_speed_heading}, so the expression \eqref{eq:r_tilde_b} is equivalent to 
\begin{align}
    \dot{\tilde{r}}
    &=\cos\tilde{\psi}(u_x\cos\theta + u_y\sin\theta) - \sin\tilde{\psi}(-u_x\sin\theta + u_y\cos\theta)-\dot{\bm{g}}\cdot\bm{e}_r- \dot{R}\;. \label{eq:r_tilde_c}
\end{align}
By transforming the coordinates from $[u_x, u_y]^{\rm T}$ to $[u_r, u_\theta]^{\rm T}$, using \eqref{eq:coordinate_rotation}, the radial error rate is
\begin{align}
    \dot{\tilde{r}}&= u_r\cos\tilde{\psi} - u_\theta \sin\tilde{\psi} - \dot{\bm{g}}\cdot\bm{e}_r- \dot{R}\;.\label{eq:radius_rate_error}
\end{align}
By using $\tilde{r} = r - R$ and $\tilde{\psi}=q_\psi - \psi_d$ with the steering controller \eqref{eq:controller_proportional} and substituting the result with \eqref{eq:radius_rate_error} into \eqref{eq:lyapunov_dot},
\begin{align}
    \dot{V}_2 
    &= 
    \frac{ (u_r\cos\tilde{\psi} - u_\theta \sin\tilde{\psi} - \dot{\bm{g}}\cdot\bm{e}_r- \dot{R}) \beta\frac{2}{\pi}\arctan{(\beta\tilde{r})}}{1 + \beta^2\tilde{r}^2}\notag \\
    & + \tilde{\psi}  \left\{ k_\psi\left(\psi_d - q_\psi\right) + \dot{\psi}_d  - \dot{\psi}_d
     +\left(\left(\bm{g}\cdot\bm{e}_r + \dot{R}\right)\frac{1-\cos(q_\psi - \psi_d)}{q_\psi - \psi_d}\right. 
    \left.+ \frac{u_\theta\sin(q_\psi - \psi_d)}{q_\psi - \psi_d}\right)\frac{\beta\frac{2}{\pi}\arctan\left(\beta(r - R)\right)}{1 + \beta^2(r - R)^2} \right \} 
    \\
    & = -k_\psi \tilde{\psi}^2 + \tilde{\psi}\left(\frac{u_\theta \sin\tilde{\psi}}{\tilde{\psi}} + \frac{(\bm{g}\cdot\bm{e}_r + \dot{R})(1 -\cos{\tilde{\psi}})}{\tilde{\psi}}\right)\frac{\beta\frac{2}{\pi}\arctan{\beta\tilde{r}}}{1 + \beta^2\tilde{r}^2} \notag\\
    & -\left(u_\theta \sin\tilde{\psi} + \dot{\bm{g}}\cdot\bm{e}_r+ \dot{R}- u_r\cos\tilde{\psi}\right)\frac{\beta\frac{2}{\pi}\arctan{(\beta\tilde{r})}}{1 + \beta^2\tilde{r}^2}\;.
\end{align}
Simplifying and canceling like terms,
\begin{equation}
    \dot{V}_2 = -k_\psi \tilde{\psi}^2 - \left((u_r - \dot{\bm{g}}\cdot\bm{e}_r - \dot{R})\cos\tilde{\psi}\right)\frac{\beta\frac{2}{\pi}\arctan{\beta\tilde{r}}}{1 + \beta^2\tilde{r}^2}\;.\label{eq:simplifed_lyapunov_field}
\end{equation}
Substituting $u_r$ from \eqref{eq:controller_expanded} into \eqref{eq:simplifed_lyapunov_field},
\begin{align}
    \dot{V}_2 
    & = -k_\psi \tilde{\psi}^2 -\frac{\left(v - v_g -\left|\dot{R}\right|\right)\cos(\tilde{\psi})\beta\frac{4}{\pi^2}\arctan^2{\beta\tilde{r}}}{1 + \beta^2\tilde{r}^2}\;.
    \label{eq:lyapunov_dot_expanded}
\end{align}
Because the orbit satisfies Theorem~\ref{thm:orbit}, Lemma~\ref{lem:UAV_speed_varying} is also satisfied and $\dot{V}_2 < 0$ for $k_\psi > 0$, $\beta > 0$, $r > 0$, $-\frac{\pi}{2} \leq \tilde{\psi} \leq \frac{\pi}{2}$, and all $\tilde{r}$ where $\tilde{\psi}$ and $\tilde{r}$ are non-zero. 
For the system to attract to the orbit $O(\theta, t)$ for all $\tilde{\psi} \in \mathbb{S}$ the inequality $\dot V_2 < 0$ from \eqref{eq:lyapunov_dot_expanded}
must also be satisfied on the intervals $[-\pi, -\frac{\pi}{2}]$ and $[\frac{\pi}{2}, \pi]$.  Let
\begin{equation}
    \epsilon = \frac{\arctan^2\beta\tilde{r}}{1 + \beta^2\tilde{r}^2}\;
    \label{eq:arctan_max}
\end{equation}
denote the coefficient appearing in  \eqref{eq:lyapunov_dot_expanded}. The bounds of $\epsilon$ were found using 
to be $0 \leq \epsilon \leq\epsilon_{\rm max}$ where $\epsilon_{\rm max} $ is the solution to the equation $1 - x\arctan{x}=0$, which is approximately $\epsilon_{\rm max} \approx 0.31483$. The function $\epsilon$ can be rewritten as
\begin{equation}
    \epsilon(x)=\frac{\arctan^2{x}}{1 + x^2}\;,\notag
\end{equation}
where $x=\beta\tilde{r}$. To find the extrema of $\epsilon(x)$, find where its derivative is zero:
\begin{equation}
    \frac{{\rm d} \epsilon}{{\rm d}x}= -\frac{2\arctan x \left(x \arctan x-1\right)}{\left( 1 + x^2\right)^2}=0\;.
    \label{eq:partial_x}
\end{equation}
The derivative is zero when $x=0$ and
\begin{equation}
    x^*\arctan x^* - 1 =0\;,
    \label{eq:darctanx}
\end{equation}
resulting in local extrema $\epsilon(0)=0$ and $\epsilon(x^*)\approx 0.31483$.
The value of $\epsilon$ as $x$ approaches $\infty$ or $-\infty$ by L'Hopital's rule is
\begin{align}
        \lim_{x\rightarrow \pm\infty}\epsilon(x) & =\lim_{x\rightarrow\pm\infty}\frac{x\arctan{x}}{1 + x^2}
        \notag\\
        &=\lim_{x\rightarrow \pm\infty}\frac{1}{2(1 + x^2) + 4x^2} + \frac{1}{2(1 + x^2)}
        =0\;.\notag
\end{align}
Thus, the coefficient appearing in \eqref{eq:lyapunov_dot_expanded}  is bounded by
\begin{equation}
    0 < \frac{\frac{4}{\pi^2}\beta\arctan^2\beta\tilde{r}}{1 + \beta^2\tilde{r}^2} \leq\frac{4}{\pi^2}\beta\epsilon_{\rm max}\;.
    \label{eq:arctanbeta_bounds}
\end{equation}
To find a maximum of \eqref{eq:lyapunov_dot_expanded} with respect to $\tilde{\psi}$ for the ranges $[-\pi, -\frac{\pi}{2}]$ and $[\frac{\pi}{2}, \pi]$ find the value on the bounds, and at local extrema.
At the bounds, substituting $\tilde{\psi}=\pm \pi$ into \eqref{eq:lyapunov_dot_expanded}, the inequality $\dot V < 0$ leads to
\begin{equation}
     k_\psi > v \beta\frac{4}{\pi^4}\epsilon_{\rm max} \geq \frac{\left(v - v_g -\left|\dot{R}\right|\right)\beta\frac{4}{\pi^4}\arctan^2{\beta\tilde{r}}}{1 + \beta^2\tilde{r}^2}\;.
     \label{eq:lyapunov_extrema_region_bounds}
\end{equation}
Substituting $\tilde{\psi}=\pm\frac{\pi}{2}$ into \eqref{eq:lyapunov_dot_expanded}, the inequality $\dot V < 0$ leads to $
    k_\psi > 0$. 
Local extrema occur where the derivative is zero:
\begin{align}
    \frac{\partial}{\partial \tilde{\psi}}\dot{V}_2  =
    0 & = -2k_\psi\tilde{\psi} + \frac{\left(v - v_g -\left|\dot{R}\right|\right)\sin{\tilde{\psi}}\beta\frac{4}{\pi^2}\arctan^2{\beta\tilde{r}}}{1 + \beta^2\tilde{r}^2}\notag\\
    \frac{\sin{\tilde{\psi}}}{\tilde{\psi}} & = \frac{2k_\psi\left(1 + \beta^2\tilde{r}^2\right)}{\left(v - v_g -\left|\dot{R}\right|\right)\beta\frac{4}{\pi^2}\arctan^2{\beta\tilde{r}}}\label{eq:partial_psi_zero}\;.   
\end{align}
The function $\frac{\sin{\tilde{\psi}}}{\tilde{\psi}}$ has a range of $[0, 1]$ over the domain $[-\pi, \pi]$ and is equal to zero at $\{\pi,-\pi\}$. The term on the right is independent of $\tilde{\psi}$. The function $\dot{V}_2$ may have local extrema on the bounds $\tilde{\psi} \in  [-\pi, \pi]$ if 
\begin{equation}
    0 \leq k_\psi \leq  \left(v - v_g -\left|\dot{R}\right|\right)\beta\frac{2}{\pi^2}\frac{\arctan^2\beta\tilde{r}}{1 + \beta^2r^2} < v\beta\frac{2}{\pi^2}\epsilon_{\rm max}\;,
    \label{eq:lyapunov_extrema_gain_bound}
\end{equation}
since this results in a value of $k_{\psi}$ where the value of the right-hand-side \eqref{eq:partial_psi_zero} is in the range $[0, 1]$.
The opposite of the upper bound in \eqref{eq:lyapunov_extrema_gain_bound} results in a value of the right-hand-side of \eqref{eq:partial_psi_zero} outside the range $[0, 1]$; Therefore, avoiding local extrema in the regions $\tilde{\psi} \in [-\pi, \pi]$, which includes the regions $\tilde{\psi}\in[-\pi, -\pi/2]$ and $\tilde{\psi}\in[\pi/2,\pi]$. For a region with no local extrema, the region's boundary contains the maximum and minimum values of a function.
The value of $k_\psi$ needed to avoid local extrema is \eqref{eq:gain_bounds}, greater than the bound \eqref{eq:lyapunov_extrema_region_bounds}. As a result, the bound on $k_\psi$ for the negative definiteness of \eqref{eq:lyapunov_dot_expanded} is \eqref{eq:gain_bounds}. A negative definite \eqref{eq:lyapunov_dot_expanded} implies that $V_2(t) \leq V_2(0)$ and that $\tilde{\psi}$ and $\tilde{r}$ are bounded. 

If $\frac{\rm d}{{\rm d}t}\dot V_2$ is bounded then $\dot{V}_2$ is uniformly continuous. By expanding,
\begin{align}
    \frac{\rm d}{{\rm d}t}\dot{V}_2  = \ddot{V}_2  & =-2k_{\psi}\tilde{\psi}(u_\psi - \psi_d) - \frac{(u_{\psi} - \psi_d)(v - v_g - |\dot{R}|)\cos{\tilde{\psi}}\beta\frac{4}{\pi^2}\arctan^2\beta\tilde{r}}{1 + \beta^2\tilde{r}^2}\notag\\ &-\frac{\beta(v - v_g - |\dot{R}|)\left(u_r\sin{\tilde{\psi}} - u_\theta\cos\tilde{\psi} - \bm{g}\cdot\bm{e}_r - \dot{R}\right)\sin{\tilde{\psi}}\beta\frac{8}{\pi^2}\arctan\beta\tilde{r}}{1 + \beta^2\tilde{r}^2}\notag\\
    &+\frac{\beta^2(v - v_g - |\dot{R}|)\left(u_r\sin{\tilde{\psi}} - u_\theta\cos\tilde{\psi} - \bm{g}\cdot\bm{e}_r - \dot{R}\right)\tilde{r}\sin{\tilde{\psi}}\beta\frac{8}{\pi^2}\arctan^2\beta\tilde{r}}{(1 + \beta^2\tilde{r}^2)^2}\;,
    \label{eq:lyapunov_ddot}
\end{align}
the boundedness can be checked. The function \eqref{eq:lyapunov_ddot} is bounded because $\tilde{\psi}$, $\tilde{r}$ , $v_g$, $\dot{R}$, $\frac{\arctan^2(\beta \tilde{r})}{1 + \beta^2\tilde{r}^2}$, and $u_\psi$ are bounded. The steering controller $u_{\psi}$ \eqref{eq:controller_proportional} is bounded locally around $|\tilde{r}| \leq \delta$ for some $0 < \delta < R $ because $\frac{1-\cos{x}}{x}$, and $\frac{\sin{x}}{x}$ are bounded, $\dot{\psi}_d$ is bounded locally around $|\tilde{r}| \leq \delta$, and $||\bm{u}|| = v$ is a constant. (The value of $\dot\psi_d$ has a singularity at $r=0$, but it is locally bounded around the orbit $O$.)
Because $\ddot{V}_2$ is bounded, rate $\dot{V}_2$ is uniformly continuous. Thus, by Corollary~\ref{cor:barbalts}, the rate $\dot{V}_2 \rightarrow 0$ as $t\rightarrow\infty$ locally around the orbit $O$.

If the system \eqref{eq:dynamics} uses the steering controller \eqref{eq:controller_proportional} with gain constraint \eqref{eq:gain_bounds}, the scalar function \eqref{eq:lyapunov} $V_2 > 0$ for all $[\tilde{\psi}, \tilde{r}]^{\rm T} \neq \bm{0}$, the rate \eqref{eq:lyapunov_dot} $\dot{V}_2 < 0$ is negative definite except when $[\tilde{\psi}, \tilde{r}]^{\rm T} = \bm{0}$, and the rate $\dot{V}_2 \rightarrow 0$ as $t\rightarrow\infty$ locally around the orbit $O$. The value of the scalar function $V_2$ decreases along the trajectory of the system \eqref{eq:dynamics} to a lower bound. That lower bound is when $[\tilde{\psi}, \tilde{r}]^{\rm T} = \bm{0}$
because $\dot{V}_2 \rightarrow 0$ as $t\rightarrow\infty$, and the rate $\dot{V}_2$ is only zero along the orbit $O$.
Therefore, the system \eqref{eq:dynamics} is locally asymptotically 
attracted to the orbit $O$ when \eqref{eq:gain_bounds} is satisfied.
    \end{proof}
\end{lem}
\begin{remark}
    \label{rem:inside_orbit}
    If the system \eqref{eq:dynamics} enters the inside of the orbit $r \leq  R$ then 
    it may encounter a singularity at $r=0$. The singularity violates Corollary~\ref{cor:barbalts} in the set $r \leq R$ because the controller $u_\psi$ is unbounded, causing $\dot V_2$ to not be uniformly continuous. However, if the singularity is excluded and the set $r \geq \tau R$ is considered, for a choice of $0<\tau \leq 1$, then the controller $u_\psi$ is asymptotically attracted to the orbit in this set. The control $u_\psi$ has a local  maximum  along $r=\tau R$. Thus, the system \eqref{eq:dynamics} is asymptotically attracted to the orbit $O$ for some $r\geq \tau R$.
\end{remark}

\subsection{Controller Turning}
\begin{figure}[h!]
    \centering
    \includegraphics[width=.3\textwidth]{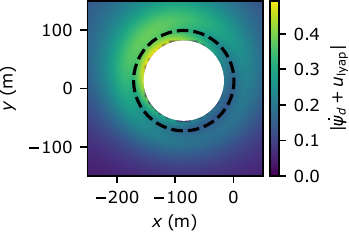}
    \caption{Tuning the $\beta$ parameter pertaining to attracting to the orbit using a grid search. }
    \label{fig:beta_tune}
\end{figure}

From Lemma~\ref{lem:controller} and Remark~\ref{rem:inside_orbit}, the steering controller \eqref{eq:controller_proportional} is bounded for all $r \geq \tau R$.
The term
\begin{equation}
    |u_{\rm lyap}| = \left|\left(\left(\dot{\bm{g}}\cdot\bm{e}_r + \dot{R}\right)\frac{1-\cos(q_\psi - \psi_d)}{q_\psi - \psi_d} + \frac{u_\theta\sin(q_\psi - \psi_d)}{q_\psi - \psi_d}\right)\frac{\beta\frac{2}{\pi}\arctan\left(\beta(r - R)\right)}{1 + \beta^2(r - R)^2}\right|< v\beta\frac{2}{\pi}\lambda_{\rm max}\;
    \label{eq:control_stability_term}
\end{equation}
is upper bounded for all possible $r \geq \tau R$ and $q_\psi \in \mathbb{S}$ because the term inside the parenthesis in \eqref{eq:control_stability_term} is at most $v$, and the term outside the parenthesis is bounded $-\lambda_{\rm max} \leq \frac{\arctan\left(\beta(r - R)\right)}{1 + \beta^2(r - R)^2} \leq \lambda_{\rm max}$ similar to \eqref{eq:arctan_max}. The approximate maximum is $\lambda_{\rm max}\approx0.41195$ and occurs when $\beta\tilde{r}\arctan(\beta \tilde{r}) - \frac{1}{2}=0$. The parameter $\beta$ controls the magnitude of both $\dot{\psi}_d$ and $u_{\rm lyap}$, and is tuned so the controller stays within the limits \eqref{eq:dynamics_constraint}. A grid search the value of $\beta$ can be seen in Fig.~\ref{fig:beta_tune} where the POI's speed is $v_g = 5$~m/s, the UAV's speed is $v = 20$~m/s, the UAV's turn radius is $r_{\rm min} = 50$~m, the radius of the orbit is $R = 85.9$~m, the radius rate of change is $\dot{R} = 1.30$~m/s, and the tuning parameter $\beta = 0.025$. The maximum value found by the grid search for $r \geq R$ was $|\dot{\psi}_d + u_{\rm lyap}| = 0.378$~rad/s which is less than the turn rate limit of $\frac{v}{r_{\rm min}} = 0.4$~rad/s. 
Therefore, the control law \eqref{eq:controller_proportional} is feasible for the system \eqref{eq:dynamics} with constraint \eqref{eq:dynamics_constraint} for this particular set of parameters $r_{\rm min}$, $R$, $\dot{R}$, $\beta$. Repeating this process for other $R$, $\dot{R}$ values along the locus of orbits $O$ can ensure choice of parameters yields a controller than can attract the system $\eqref{eq:dynamics}$ with constraint \eqref{eq:dynamics_constraint} to the locus of orbits for all time $t \in [t_0, t_f]$. If the system \eqref{eq:dynamics} enters the inside of the orbit $r\leq R$ then the terms $u_{\rm lyap}$, and $ \dot\psi_d$ may saturate the controller $u_\psi$ causing system \eqref{eq:dynamics} to be unable to follow a trajectory of \eqref{eq:controller_expanded}. However, with a choice of $\beta$ and $\tau$ the quantity $|u_{\rm layp} + \dot\psi_d|<u_{\rm max}$ for $r \geq \tau R$ where $0<\tau\leq 1$  (from Remark~\ref{rem:inside_orbit}), the controller is asymptotically attractive for $r\geq \tau R$ considering \eqref{eq:dynamics_constraint}.

\section{Numerical Study Results}
\label{sec:results}

This section presents numerical results from the reference orbit design strategy of Sec.~\ref{sec:polygons} and feedback controller of Sec.~\ref{sec:feed_controller}\footnote{The implementation of this study can be found at \url{https://github.com/robotics-uncc/DubinsVisibilityTracking}.}
The environment is modeled after the city-center of Charlotte, North Carolina with building geometry from Open Street Map \cite{OpenStreetMap} extracted from a bounding box,
$( 80.8471^{\circ}$W, $35.2209^{\circ}$N) and $(80.8335^{\circ}$W, $35.2289^{\circ}$N).
The POI moves along the road northeast at constant velocity $v_{g} = 5$~m/s from $\bm{g}=[-150, -45]^{\rm T}$, then takes a turn and continues southeast along the next road. The POI takes $48.16$~s to reach the end of the second road and the UAV's motion is simulated over the same period.
\begin{figure}[t]
    \centering
    \includegraphics[width=.8\textwidth]{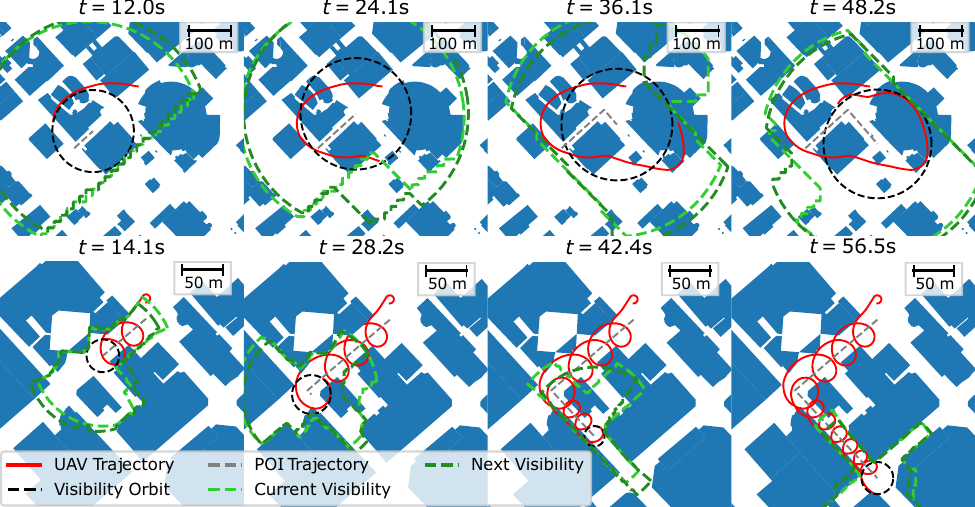}
    \caption{Two examples of a UAV (red) tracking a moving POI through an urban environment (gray).}
    \label{fig:trajectory}
\end{figure}
Two examples are presented in Fig.~\ref{fig:trajectory}---the first and top considers a higher altitude UAV with a larger turn radius, the second and bottom considers a lower altitude UAV with a tighter turn radius.

In the first example, the UAV flies with a constant speed of $v=20$~m/s at an altitude of $h_{\rm UAV}=300$m with a initial position of $\bm{q}=[0, 100, -\pi]^{\rm T}$.
 To track the moving POI, the visibility along the known path is calculated with the adaptive discretization algorithm, Sec.~\ref{sec:adaptive}, with an initial sample spacing of $||\bm{g}_i - \bm{g}_{i + }|| =20$~m and a cutoff of $d_{\rm cutoff} = 20 \times 10^6$~${\rm m}^3$. The VVs are calculated with $d_{\rm max} = 400$~m. The minimum turn radius of the UAV was set to $r_{\rm min} = 50$~m, $\beta=0.025$ was selected from Fig.~\ref{fig:beta_tune}, and the steering gain set was to $k_\psi=20$.
 The initial radius of the orbit is $R_i = 158$~m and the final radius is $R_f=127$~m. The trajectory was solved with the Runge-Kutta 45 method \cite{Dormand1980}. As seen in the two trajectories in Fig.~\ref{fig:trajectory} the UAV (red trajectories) can start at initial conditions off of the visibility orbit with a heading not aligned with the vector fields and steer onto the morphing visibility orbits. The visibility orbits (black dashed circles) from Sec.~\ref{sec:circular_orbits} are within the VVs (dashed green lines) in Fig.~\ref{fig:trajectory}, and the UAV's trajectories deviate minimally from the orbits. With the assumption that the POI remains visible during the time steps between the sampled orbits, the UAV can maintain the visibility of the POI for its trajectory (dashed gray).

\begin{figure}[t]
    \centering
    \includegraphics[width=.4\textwidth]{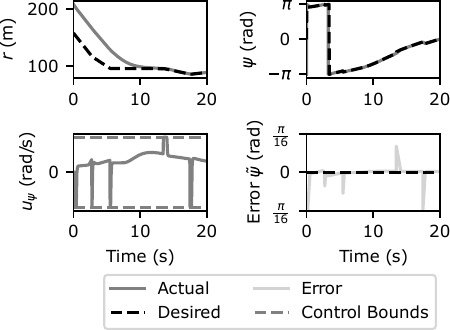}~
    \includegraphics[width=.4\textwidth]{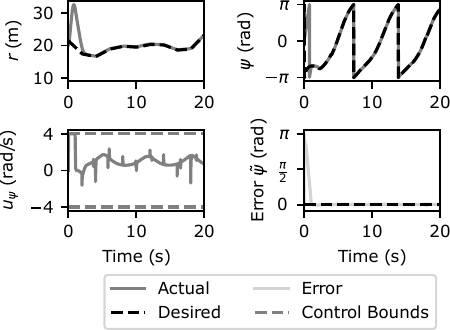}
    \caption{A plot of the error and control in the first $20$~s of the simulations (first: left and second: right). 
    }
    \label{fig:error}
\end{figure}

The radial distance to the orbit, the heading angle, and their errors are shown in Fig.~\ref{fig:error}. 
The graphs show that the proportional controller with a feed-forward term forces the UAV onto the vector field, which guides the UAV onto the time-varying orbit. The time-varying orbit $O$ is a linear piecewise continuous function with a discontinuous derivative that occurs at the orbit transition points where $\dot{\bm{g}}$ changes abruptly. This violates the assumption of \cite{Goncalves2010}, which assumes a bounded second derivative. The violation of the assumption causes the UAV to deviate from the orbit slightly at each orbit transition point. Still, the controller can quickly force the UAV back onto the orbit. This occurrence can be seen on the error graphs where the error spikes up every time the rate of change of the radius changes. The implementation of the controller considers the minimum of traveling clockwise or counter-clockwise as the heading error, allowing the UAV to turn counter-clockwise from $\pi$ to $-\pi$. The control for the UAV is shown in Fig.~\ref{fig:error}.
There are large spikes in the control effort whenever the rate of change of the radius discontinuously changes. The saturation of the controller only comes into effect in this example when the UAV's proportional controller exceeds the turn rate control limit \eqref{eq:dynamics_constraint}. 

The second example of the tracking approach is shown on the bottom of Fig.~\ref{fig:trajectory}. In this example, the UAV has a very sharp turning radius of $r_{\rm min} =5$~m (e.g., representative of a smaller more agile system), a speed of $v=20$~m/s, an altitude of $h_{\rm UAV}=50$~m, an initial configuration of $\bm{q}=[-45, 450, 0]^{\rm T}$, and a maximum sensing distance of $d_{\rm max} = 100$~m. The controller uses $\beta=0.5$ and  $k_\psi = 40$. The VVs were calculated along a different known POI trajectory that starts at $\bm{g}=[-45, 430]^{\rm T}$ and moves southwest with speed $v_g=5$~m/s. As before, the adaptive discretization algorithm in Sec.~\ref{sec:adaptive} is used to compute the VVs. An initial spacing of $||\bm{g}_i - \bm{g}_{i + 1}|| = 20$~m  and a cutoff of $d_{\rm cutoff} =500 \times 10^3$~${\rm m}^3$ are assumed. The more maneuverable UAV makes tighter turns at a lower altitude with stricter sensing constraints allowing for and orbit with an initial radius of $R_i = 21.5$~m and final radius of $R_f=19.7$~m. The UAV orbits the POI for $56.52$~s and circles the POI more times in bottom of Fig.~\ref{fig:trajectory} compared to the top Fig.~\ref{fig:trajectory}. The trajectory of the UAV is shown in red, and the attractive VO is shown in black. The UAV model quickly turns just over $180^\circ$ during the first second of the simulation onto the vector field \eqref{eq:controller_expanded} using \eqref{eq:controller_proportional}. The UAV is able to remain on the VO inside the VPs (green dashed), maintaining a view of the POI (gray dashed).
During the first $1$s of the trajectory, the control shown in Fig.~\ref{fig:error} is saturated to the maximum value given by the turn radius and speed $u_{\psi, \rm max} = v/r_{\rm min} = 4$ rad/s because the UAV needs to turn around quickly. The spikes in the controller appear as a result of the discontinuous changes in $\dot{R}$ occurring when the UAV transitions from following the locus of orbits between $R_{i - 1}$ and $R_{i}$ to the locus of orbits between $R_i$ and $R_{i + 1}$. The spikes in the plot of the control for the second simulation in Fig.~\ref{fig:error} appear smaller than the spikes in  the plot of the control for the first simulation in Fig.~\ref{fig:error} because the maximum control $u_{\psi, \rm max}$ is larger. 

An example of the error for the second simulation is shown on the right half of Fig.~\ref{fig:error}. The plots show the distance of the UAV $r$ to the orbit $O$ in the radial direction and the difference between the desired $\psi_d$ and actual heading $\psi$. The initial condition for the UAV starts inside the orbit and grows larger as the UAV turns to minimize the difference between the desired and actual headings. The increased number of orbits about the POI can also be seen in the plot of heading angle $\psi$ by the increase in the number of transitions between $\pi$ and $-\pi$. The increasing nature of the heading angle is because the controller forces CCW circulation about the POI. To circulate CW, use the negative solution of \eqref{eq:circulate}.

\section{Flight Test Experiment}
\label{sec:flight_test}

\begin{figure}[h!]
    \centering
    \includegraphics[width=3.25in]{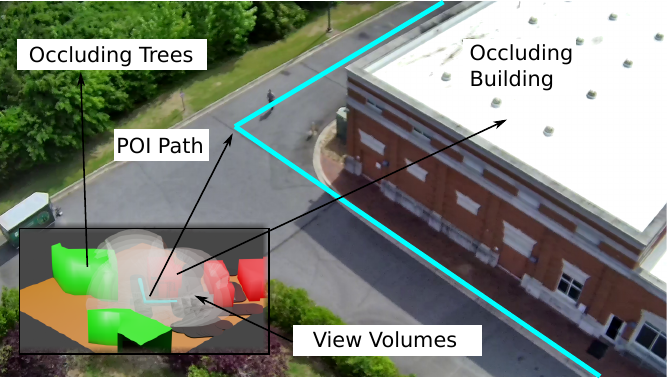}
    \caption{Model of the simulated and real-world  environment used during the experiment. 
    }
    \label{fig:setup}
\end{figure}

The proposed method was tested on April 27th, 2025, using a multicopter UAV to track a moving POI on the University of North Carolina at Charlotte campus. The urban-like environment was modeled using data from OpenStreetMap \cite{OpenStreetMap} extracted around the point         (35°18'46.3"N 80°44'24.8"W) (see  Fig.~\ref{fig:setup}). In the model of the campus, nearby buildings (height 10~m) and trees (height 30~m) may obscure the UAV's view of the POI. The experiment was performed in an augmented reality setting---a real UAV was flight tested, but a virtual gimballed camera was used to track a virtual POI in the environment. The UAV used in the experiment was a custom-built multirotor \cite{Hague2024} with an autopilot running Ardupilot \cite{Ardupilot} and a companion computer for guidance. The path of the POI was obtained prior to the experiment by collecting the GNSS trace of a Clearpath Robotics Jackal unmanned ground vehicle (UGV) steered by a human operator. The UGV's speed varied from $0.3$~m/s to $0.4$~m/s and was approximated as a couple of constant velocity sections with an instantaneous velocity change at the junction of the two path segments. To create the constant sections, two lines of best fit (cyan lines) were fitted to the corresponding GPS trace segments Fig.~\ref{fig:setup}, resulting in a path with a total length of $90.5$~m. This pre-recorded path, $\hat{\bm{g}}$, consisting of latitude and longitude positions over time, was then processed on a laptop computer using the adaptive discretization algorithm with an initial discretization of $||\bm{g}_i - \bm{g}_{i + 1}|| = 20$~m and a cutoff of $d_{\rm cutoff} = 10^5$~${\rm m}^3$. The VVs along the path were computed using a maximum sensing radius of $d_{\rm max}=50$~m shown as white volumes in Fig.~\ref{fig:setup}. The UAV was assumed to fly at a constant velocity of $v=3$~m/s at an altitude of $h_{\rm UAV}=35$~m and with a minimum turn radius of $r_{\rm min}=5$~m. The altitude was chosen with allowance so that the UAV flies over all the obstacles. The locus of orbits was created with Alg.~\ref{alg:circular}. 
To control the UAV, the onboard companion computer uses the Robot Operating System (ROS) Noetic \cite{ROS} to communicate with the UAV. 
Since the experiment was conducted with a multirotor, rather than a fixed-wing vehicle, the heading of the vehicle could be independently controlled from the velocity. As a result, the velocity from the designed vector field \eqref{eq:controller_speed_heading} (with gain $\beta=0.5$) was used directly as a setpoint command to the autopilot via a ROS node. The required POI trajectory data to execute the guidance and control strategy is the set of orbit radii, orbit origin positions, and corresponding times.

To minimize the degree to which the virtual gimballed camera must actuate (since real gimbals have actuation limits), the UAV's yaw is controlled so the relative angle to the virtual POI is consistent with the guidance vector field (which tends to face the future position of the POI). The same ROS node that controls the velocity controls the yaw rate using a proportional controller with feed-forward term, 
$
    \dot{\psi} = k_{\rm yaw}(\psi_d - \psi) + \dot{\psi}_d\;,
$
where the UAV's yaw is denoted $\psi$, $k_{\rm yaw} = 0.6$ is a positive yaw gain, the desired yaw is \eqref{eq:psi_d}, and the desired yaw rate is \eqref{eq:dot_psi_d}. 

A software-in-the-loop (SITL) simulation was created to test before flying an experiment. The simulation used ROS with Gazebo Classic and the Ardupilot SITL package. To test different sensing constraints and altitudes to find a set of parameters for the flight experiment.
The SITL simulation compared the proposed approach of using a time-varying radius $R$ from Sec.~\ref{sec:polygons} to orbiting with a constant radius. The time-varying radius $R$ had a minimum value of  $11.2$~m and a maximum value of $23.4$~m. Moreover, given the assumed maximum sensing distance of $d_{\rm max}=50$~m and the UAV's altitude $h_{\rm UAV}= 35$~m the maximum allowable orbit radius (in the absence of obstacles) is $35$~m. Therefore, the three constant-radii SITL simulations used for comparison: $11.2$~m, $23.4$~m, and $35.0$~m.

The visibility of the virtual POI during each SITL simulation was 100\%, 84.0\%, 64.0\%, and 99.3\% for orbit radii of 11~m, 23~m, 35~m, and the dynamic radius $R(t)$, respectively.  The POI's visibility from UAV was tested using ray-tracing after the mission.
From the simulations, the POI's visibility decreases as the radius increases. However, the dynamic radius maintains $99.3\%$ visibility while maximizing the standoff distance. 
An image of the experiment setup is shown in Fig.~\ref{fig:setup}.

\begin{figure}[h]
    \centering
    \includegraphics[width=.4\textwidth]{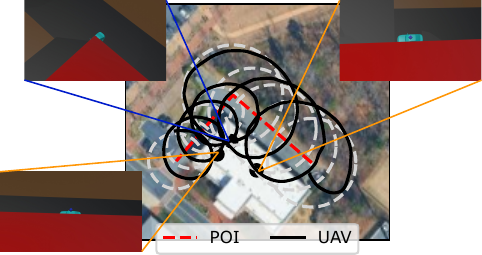}~
    \includegraphics[width=.4\textwidth]{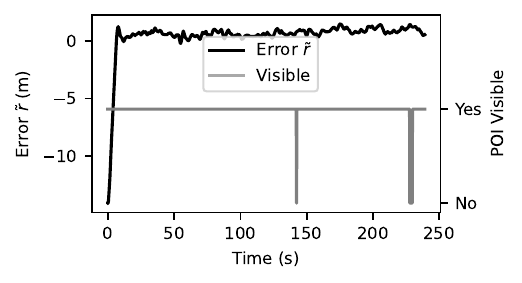}
    \caption{The trajectory, visibility, and radial error of the UAV and the POI for a visibility-informed orbit.
    }
    \label{fig:quad_flight}
\end{figure}

In the real-world flight experiment, the virtual POI moves along the cyan path in Fig.~\ref{fig:setup}, while the UAV circles above. The UAV hovers above the virtual POI until the mission is started, and both the virtual POI and real UAV start moving at the same time.
A total of six flight tests were conducted: three using the time-varying, visibility-informed $R$ orbits and three using a constant-radius orbit of $R = 35$~m to highlight the importance of considering occlusions. %
The resulting visibility percentage of the POI during each experiment is 98.0\%, 98.0\%, and 99.0\% for the VOs and 70.3\%, 71.1\%, and 66.5\% for the fixed non-visibility orbits.  The visibility was found in post-processing by ray-tracing the line between the real UAV position and the virtual POI position in the virtual environment. The UAV was able to successfully view the POI an average of 98.3\% of the time using the proposed method in the experiment; This was on average 29\% more than the fixed non-visibility method for the same tracking mission.
The overall visibility percentages are comparable to the SITL simulations. The proposed method did not produce 100\% visibility due to tracking errors. 
Figure~\ref{fig:quad_flight} provides a more detailed time-history of the trajectory and radial error for one of the experiments using the visibility-informed orbit. 
The left panel of Fig.~\ref{fig:quad_flight} shows the  trajectory of the UAV continuously circling the virtual POI. Selected images from the virtual UAV's camera sensor are shown. Two of the images show the center point of the POI visible (rendered as a dark blue cube resting on top of a cyan car), while another shows the center point of the POI obscured by the corner of the red building.
The VO, show as gray dashed lines, start large because the UAV is far from the trees and buildings (approximately 7.8~m from the building and 24.3~m from the trees), and then shrink once the UGV turns the corner because the UGV is closer to the building (approximately 3.2~m from the building and 34.3~m from the trees).
The radial error in the right panel of Fig.~\ref{fig:quad_flight} shows a mean deviation of $0.72$~m with a maximum of $1.77$~m and a minimum of $-0.23$~m after converging to the circular orbit during the experiment. 

\begin{figure}[h]
    \centering
    \includegraphics[width=.4\textwidth]{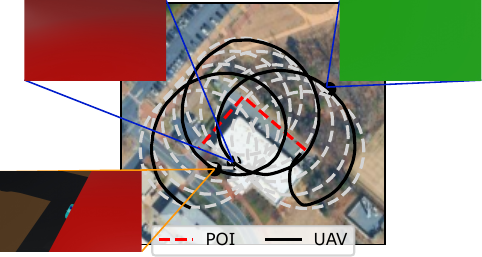}~
    \includegraphics[width=.4\textwidth]{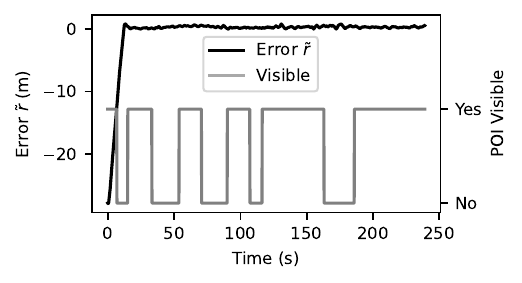}
    \caption{The trajectory, visibility, and radial error of the UAV and the POI for a non-visibility-informed orbit.
    }
    \label{fig:quad_flight_poor}
\end{figure}

Lastly, Fig.~\ref{fig:quad_flight_poor} shows results from an experiment using the fixed orbit radius wherein the visibility was 66.5\%.
Two of the images in the left panel of Fig.~\ref{fig:quad_flight_poor} show the POI obscured by the trees (green) and building (red). The mean radial error after converging to the constant radius orbit is $0.29$~m with a maximum value of $0.81$~m and a minimum value of $-0.22$~m. Both experiments have non-zero mean radial error. This can be explained by the controller sending fixed velocity setpoints at 10 Hz based on the estimated position rather than the continuous control signal in Sec.~\ref{sec:results}. This results in a tendency to err on the outside, rather than inside, of the orbit.

\section{Conclusion}
\label{sec:conclusion}
This work describes an approach to track a ground POI moving along a known trajectory through an urban environment along a translating circular orbit with a time-varying radius that is inscribed inside the POI's dynamic visibility volume (VV). VVs are created by adaptively discretizing points along the POI's trajectory while satisfying a constraint that ensures the visibility do not change too rapidly. Conditions are derived that guarantee the resulting orbits are feasible for the Dubins vehicle. A series of linearly interpolated orbits are then created inside the discretized visibility to approximate a continuous visibility orbit. A Lyapunov-based control design is proposed to steer the UAV onto a vector field that guides the UAV onto the visibility orbit. The approach is demonstrated through several numerical examples and through a real-world multirotor UAV tracking experiment wherein a virtual POI is tracked in-between buildings and trees on a university campus. The flight test showed that the UAV maintained the POI within line-of-sight an average of 29\% more often in comparison to a non-adaptive approach with a constant maximum-radius orbit for the chosen environment model. Limitations of the proposed approach are that both the environment and POI motion are assumed to be known a priori. Future work may consider modifying the approach to allow for an estimated POI position and uncertain building geometry. Expanding the orbits from circles to arbitrary time-varying polar curves can also enable larger standoff distances that better approximate the visibility polygons.

\bibliography{main.bib}             %

\end{document}